\documentclass[11pt]{article}

\usepackage{amsmath, amssymb, wasysym, slashed, multirow,hyperref}
\usepackage{enumerate}
\usepackage{pdfpages}
\usepackage{cite}
\usepackage{times}

\newenvironment{sciabstract}{%
\begin{quote} }
{\end{quote}}

\newcounter{lastnote}

\title{Resummation in QFT with Meijer G-functions}

\author
{Oleg Antipin,$^{1\star}$Alessio Maiezza,$^{1\ast}$  Juan Carlos Vasquez$^{2\dagger}$\\
\\
\normalsize{$^{1}$Ruder Bo\v skovi\'c Institute, Division of Theoretical Physics, Bijeni\v cka cesta 54, 10000, Zagreb, Croatia,}\\
\normalsize{$^{2}$Universidad T\'ecnica Federico Santa Mar\'ia $\&$ CCTVal, Valpara\'iso, Chile}\\
\\
\small{ E-mail: oantipin@irb.hr$^{\ast}$, amaiezza@irb.hr$^{\ast}$, juan.vasquezcar@usm.cl$^{\dagger}$}
}

\date{}

\begin{document}


\baselineskip16pt 


\maketitle


\begin{sciabstract}
We employ a recent resummation method to deal with divergent series, based on the Meijer G-function, which gives access to the non-perturbative regime of any QFT from the first few known coefficients in the perturbative expansion. Using this technique, we consider in detail the $\phi^4$ model where we estimate the non-perturbative $\beta-$function and prove that its asymptotic behavior correctly reproduces instantonic effects calculated using semiclassical methods. After reviewing the emergence of the
renormalons in this theory, we also speculate on how one can resum them. Finally, we resum the non-perturbative $\beta-$function of abelian and non-abelian gauge-fermion theories and analyze the behavior of these theories as a function of the number of fermion flavors. While in the former no fixed points are found, in the latter, a richer phase diagram is uncovered and illustrated by the regions of confinement, large-distance conformality, and asymptotic safety.
\end{sciabstract}

\section{Introduction}

The perturbative expansion in  QFT has zero radius of convergence~\cite{Dyson:1952tj} and the truncated series is strictly valid only for infinitesimal couplings. A resummation procedure is therefore required and the problem is usually approached through the so-called resurgent analysis~\cite{Resurgence}. Within this approach, one first applies the Borel-transform to the original power series by dividing the $n$-th term in the expansion by a $n!$ and thus improving the convergence. Then this improved result is analytically-continued and finally converted back to a convergent result via the Laplace transform.  Unfortunately,   the Borel-transformed series may have poles anywhere in the complex  plane that limits the radius of convergence and there are at least two known sources for these poles, namely:
\begin{enumerate}[i]
\item \emph{instantons}, classical solutions of the equations of motion, which can be traced back to the $n!$ number of Feynman diagrams at the $n$-th order of perturbation theory;
\item \emph{renormalons}, related to the Feynman diagrams of specific topology, for which the {\it finite part} grows factorially with the order of perturbation theory.

 \end{enumerate}
A well-known method to perform the analytic continuation is through the Pad\'e \emph{approximants} and the  whole resummation approach is then often called
Borel-Pad\'e resummation. It should be stressed, however, that a number of alternatives exist, as for
example the large-coupling-expansion which builds a power series expansion in the inverse of the coupling (see~\cite{Kleinert:2001ax,Aniceto:2018bis} for reviews).

In this article, we focus on a recent method of Borel-hypergeometric resummation proposed in Ref.~\cite{PhysRevB.94.165429,Mera:2018qte}, in which the Pad\'e  approximants are replaced by the more sophisticated hypergeometric functions, and the resummed  result admits a representation in terms of Meijer G-functions (MGs). The approach may be able to accurately resum divergent series with only a first few known coefficients even in the presence of instantons. Strictly speaking, also the Borel-Pad\'e resummation can do it but with the replacement of branch cuts by a string of poles and the necessity of knowing a large number of coefficients~\cite{Mera:2018qte}. Furthermore, in this work we will argue that this MG algorithm might help even to resum the renormalon series, thus alleviating this tough non-perturbative issue. Altogether, the Borel-hypergeometric  resummation provides a continuation from perturbative to non-perturbative physics (see also Refs.~\cite{Mera:2014sfa}). Although it is a powerful mathematical tool, one still has to worry which non-perturbative effects are being resummed  as genuine non-perturbative physics cannot probably be understood without a non-perturbative formulation (renormalization) of QFT.

The resummed series in terms of Meijer G-functions can shed light on another question in QFT, namely, the understanding of the renormalization group (RG) flow in the theory space of couplings. This is a fundamental task that is again obscured by a partial knowledge of the $\beta-$functions in the form of the divergent truncated series. Since the $\beta-$functions are obtained from the divergent part of the Feynman diagrams, they are free of renormalons. However, they still contain non-perturbative instantonic corrections.
Using the MG method,  we will estimate the $\beta-$functions non-perturbatively, showing that the method captures instantonic corrections that agree with the known theoretical predictions from semiclassical asymptotic expressions. Therefore, the method allows making reliable predictions on the behavior of a theory at all
energies. This is especially important for the ultraviolet (UV) completion of the theory,  which is the crux
of the definition of a fundamental theory~\cite{Wilson:1973jj,Parisi:1977}, that is,  the requirement that the RG flow is analytic at all energies. Specifically, we will search for possible non-trivial ultraviolet fixed points (UVFPs) of the non-perturbative $\beta-$function important for the notion of asymptotic safety. This notion was first introduced as a non-perturbative renormalizability condition for quantum gravity~\cite{Weinberg:1976xy} and is also central in the spontaneously-broken-local-conformal-invariance framework proposed in Ref.~\cite{tHooft:2011aa}.

While recalling the main features of the Borel-hypergeometric  resummation, in Sec.~\ref{Sec:phi} we deal with the $\phi^4$ model and its $\beta-$function. The $\phi^4$ model is perhaps the most studied QFT and thus represents an ideal benchmark for the MG algorithm.
In Subsection.~\ref{Subsec:ren}, we exploit this model to recall and stress about the emergence of renormalons and conjecture that the MG algorithm might be
used to resum them. Then in Sec.~\ref{Sec:gauge} we turn our attention
to the gauge theories, searching for a fixed point  (FP) as a function of the number of massless fermion flavors.
We draw our conclusions in Sec.~\ref{Sec:end}, and finally a technical appendix~\ref{App_MeijerG} avails to recall the MG algorithm as well as to deepen the understanding of the whole approach.

\section{The $\phi^4$ model}\label{Sec:phi}

The usual resummation procedure~\cite{Resurgence} applied to a  given divergent power series $\sum _n a_n x^n$ is formally sketched as
\begin{align}\label{sketch}
\sum _n a_n x^n \mapsto (\text{Borel transform}\Rightarrow &\text{analytic continuation in the Borel plane}\Rightarrow \nonumber \\ &  \text{Laplace transform})\,.
\end{align}
Once a large enough number of terms of the divergent series are known, the conventional Borel-Pad\'e resummation may be applied to cover the first two steps in Eq.~\eqref{sketch}. However, to be sufficiently accurate beyond the weak coupling regime,
this method needs as an input many orders of the perturbative expansion and separate summation for different branches. Here we attempt to resum the $\beta-$function in the $\phi^4$ model, by using the recent approach proposed in Ref.~\cite{Mera:2018qte}, in which the authors replace the Borel-Pad\'e approximants
with the hypergeometric and Meijer G-functions. Thanks to the flexibility of the special Meijer's G-functions, i.e. the feature to contain most of the known
special functions as particular cases, one may be able to accurately resum divergent series from the knowledge of only the first few
terms in the power expansion. Therefore, the method is claimed to be fast and enables us to analytically continue the perturbative series to the non-perturbative regime.
Hence, in the rest of this article we replace the procedure in~\eqref{sketch} with
\begin{equation}\label{MGdef}
\sum _n a_n x^n \mapsto MG\left(\sum _n a_n x^n\right) \,.
\end{equation}
where $MG$ formally symbolizes the algorithm proposed in Ref.~\cite{Mera:2018qte} and discussed in App.~\ref{App_MeijerG}.

The Borel-hypergeometric  resummation is formidably tested in the 0-dimensional functional (or partition function)~\cite{Mera:2018qte}, a regular one dimensional integral that can be solved both \emph{exactly} as a Bessel function and \emph{approximated perturbatively} for small couplings (see for instance a nice discussion in Ref.~\cite{Flory:2012nk}). In the latter approach, one finds that the perturbative series contains the instanton that causes the departure from the
exact result.  Remarkably, already in the leading approximation, the Borel-hypergeometric  resummation manages to take into account this instanton divergence with a good precision. This is somehow related to the fact that the Bessel function is indeed a particular case of a Meijer G-function and therefore since the exact result belongs to the set containing the first approximants, the convergence is quick.

The 0-dimensional Green function
considered in Ref.~\cite{Mera:2018qte} is a sort of QFT in one point of space-time, thus it is natural to extend the approach to an infinite-dimension counterpart, namely
the regular $\phi^4$ model in 4D space-time (in~\cite{Mera:2018qte}, the MG algorithm was also applied to the 3D Ising model, deeply related to $\phi^4$ in 3D). It is known that 4D $\phi^4$ model does not have fixed points~\cite{Wilson:1973jj}. This emerges partially from the perturbative computation of
the $\beta-$function and, on a more general ground, shown to be true at any order in Ref.~\cite{Wilson:1973jj} using the high-temperature expansion for a statistical mechanics model on a lattice. We reassess the issue within the framework of Meijer G-functions summation.

\subsection{Borel-hypergeometric resummation of the $\beta$ function perturbative series}

 \begin{figure}
 \centering
     \includegraphics[scale=0.4]{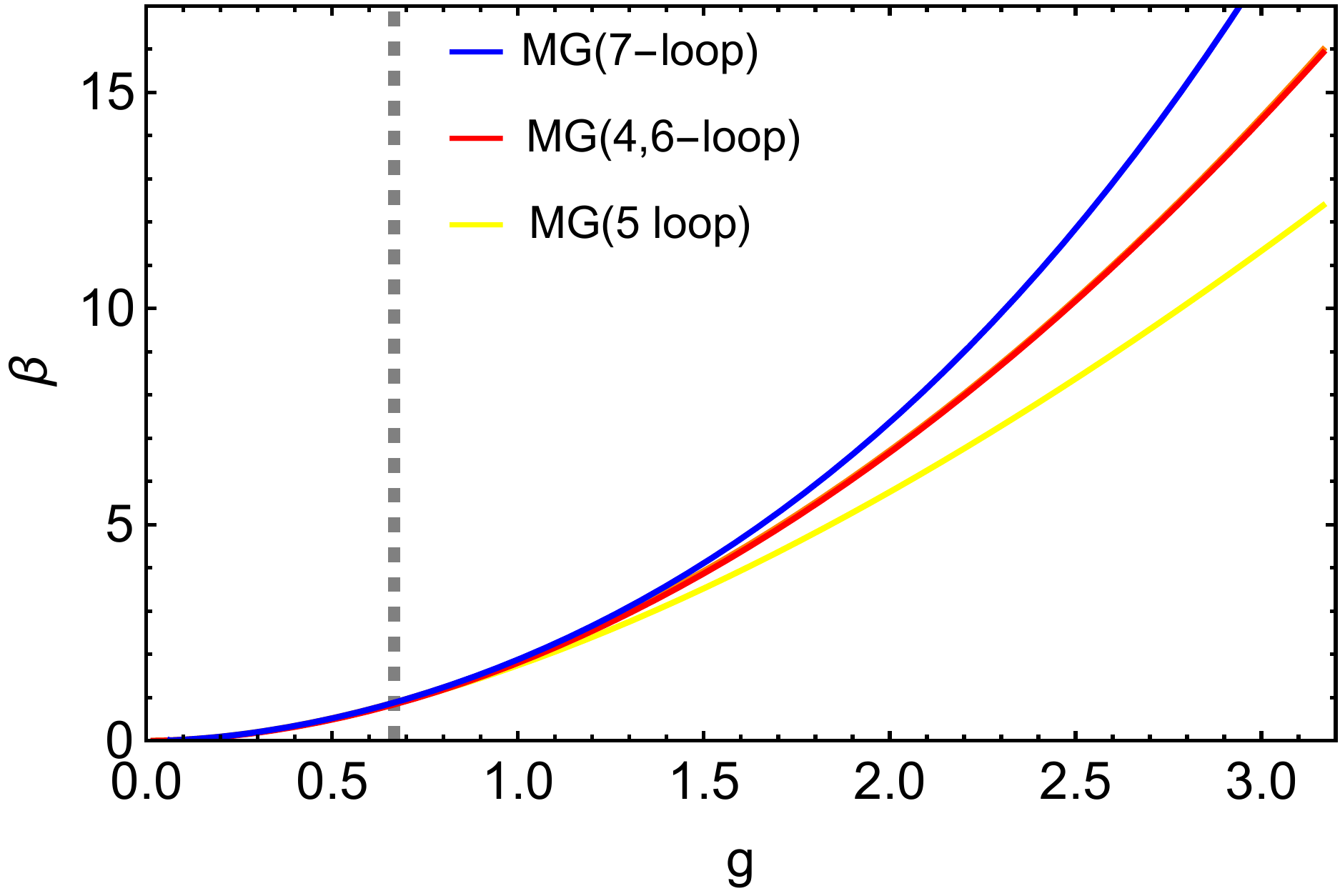}
  \caption{ Resummed  $\beta-$function as a function of $g$ from 4-7 loops input  in terms of the Meijer G-functions. The vertical dashed line signals the renormalon emerging form the finite
  part of the diagram in Fig.\ref{fig:Renormalon} and discussed in Subsec.\ref{Subsec:ren}.}
   \label{fig:phi4}
\end{figure}

The $\beta-$function of the 4D $\phi^4$ model with $\mathcal{L}_{int.}=\frac{\lambda}{4!}\phi^4$ is known up to 7-loops and in $\overline{MS}$-scheme reads~\cite{Schnetz:2016fhy}
\begin{eqnarray}\label{phi4beta}
\beta[g]&\equiv& \frac{d g}{d \lg\mu} =3 g^2-5.7 g^3+32.5 g^4 -\\ \nonumber
& & 271.6 g^5+2848.6 g^6-34776 g^7+474651 g^8 +\mathcal{O}(g^9)\,,
\end{eqnarray}
with $g\equiv\lambda/(16\pi^2)$.
We employ the algorithm described in detail in App.~\ref{App_MeijerG} to resum this $\beta-$function using as an input four, five, six and, finally, seven-loops terms known in the literature so far. From the most accurate 7-loops input, for example, we obtain:
\begin{equation}
\beta_{MG}^{(7)}[g]=3g^2 \left[1 - 10^{-15} g\, G_{3,4}^{4,1}\left(\frac{1.2}{g}|
\begin{array}{c}
 1,3.0,0.058 \\
 1,1,18.85,0.063 \\
\end{array}
\right)\right]\end{equation}
and this function, together with the lower order results is illustrated in Fig.~\ref{fig:phi4}, in which the MG summed $\beta-$function is
plotted as a function of $g$. As is clear from the figure, no fixed points emerge.

\begin{figure}
 \centering
     \includegraphics[scale=0.47]{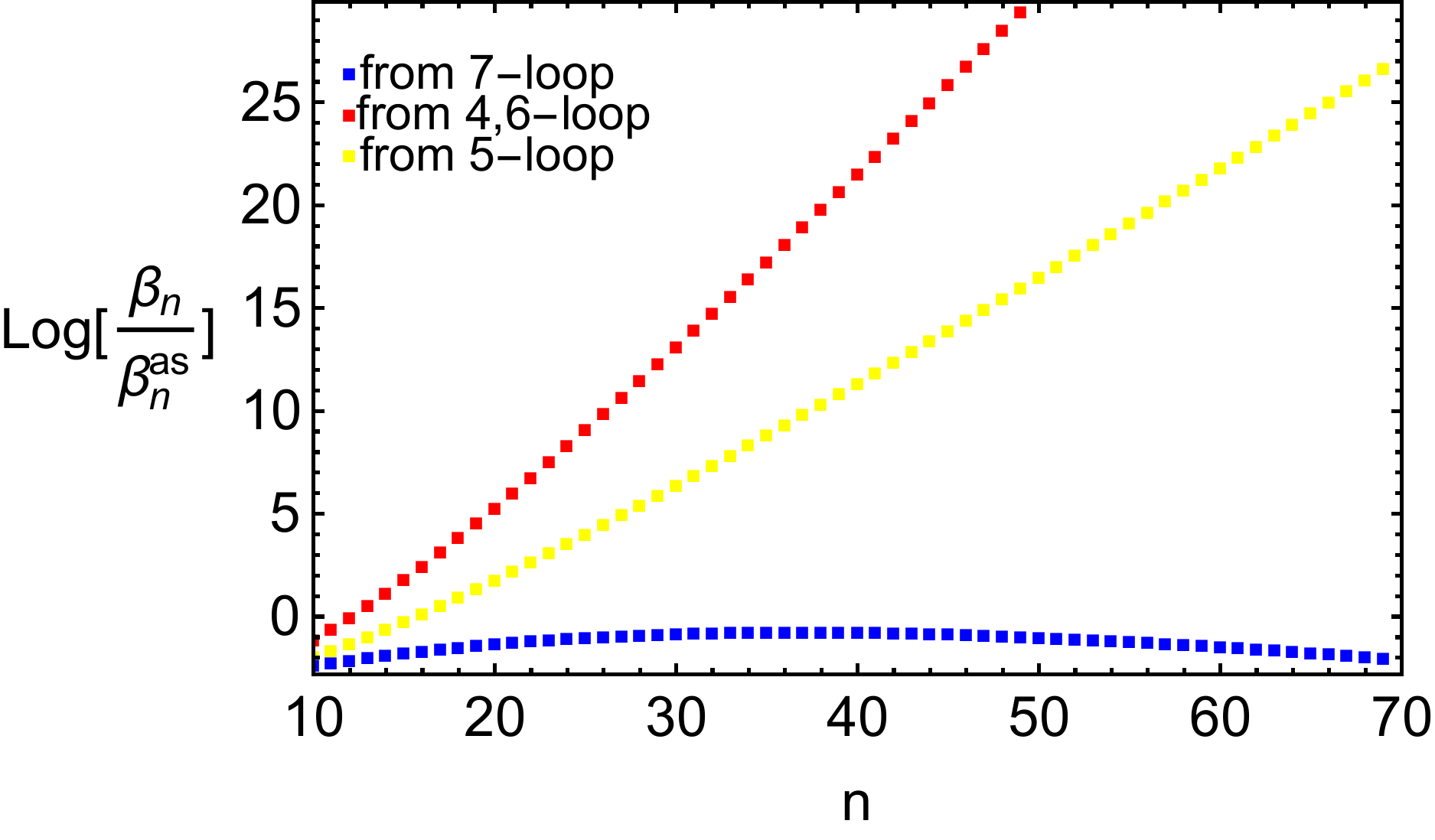}
  \caption{Large order $\beta-$function coefficients from Meijer-G at 4-7 loops normalized to their asymptotic values.}
  \label{asymptotics}
\end{figure}

The convergence of the MG approximants in Fig.~\ref{fig:phi4} is not as perfect as in the 0-dimensional case but still sufficient to capture the correct non-perturbative behavior.
It is indeed known that in this model the $\beta-$function can be represented as an asymptotic series in the
coupling $\lambda$ (or $g$), with factorially growing coefficients~\cite{Lipatov:1976ny,McKane:2018ocs}. For the $\overline{MS}$ scheme, the precise leading asymptotic behavior was first computed in~\cite{McKane:1984eq} using $4-2\epsilon$ dimensional instantons. Namely, if we denote the coefficients of the beta function by $\beta(g) =\sum_{n} \beta_n^{as} g^n$
then~\cite{Kompaniets:2017yct}
\begin{equation}
\beta_n^{as}\sim (-1)^n n! \ n^{7/2} \times const \qquad \qquad \text{as}\quad  n\to\infty\,,
\end{equation}
where $const\approx 0.024$. One can compare this asymptotic estimate with the MG summed predictions which we reexpand back in the Taylor series. In Fig.~\ref{asymptotics} we plot the logarithm of the ratio of the large order beta function coefficients $\beta_n$ as predicted by the Meijer G-function (from $n=10$ to $n=70$ loop order) to their predicted asymptotic form $\beta_n^{as}$. We notice that the result from the 7-loop MG algorithm is very close to the predicted asymptotic behavior which demonstrates that the Borel-hypergeometric resummation captures the non-perturbative instantonic effects, invisible in perturbation theory. It is worth noticing that qualitatively this is in full analogy with the 0-dimensional functional discussed above. This confirms the power of Borel-hypergeometric resummation to scan non-perturbative physics from perturbative inputs and shows the natural
applicability of the method to reconstruct the non-perturbative $\beta-$function from its truncated power series.

\subsection{Non-perturbativity vs Renormalons}\label{Subsec:ren}

So far the Reader may be induced to think that the Borel-hypergeometric resummation represents definitively a non-perturbative answer once one knows
a sufficient number of terms of a truncated series. Although the method is powerful and enables us to get reliable insights on non-perturbative physics, this may not be the case when \emph{genuine} non-perturbative effects emerge from non-perturbatively-renormalizable field theories\footnote{It is sufficient to think of realistic theory as QCD where the non-perturbativity is accompanied by chiral symmetry breaking and non-perturbative objects such as the quark condensate.}. In principle, these effects would be well described by a fundamental non-perturbative formulation of a QFT, but probably they cannot be extrapolated by means of an analytic continuation from the perturbative regime. In other words, we argue that there may be a sort of qualitative
discontinuity between perturbative and non-perturbative physics. Such discontinuity should be related with the failure of perturbative-renormalizability in the strict sense
and, in practice, might be signaled by the emergence of the renormalons~\cite{tHooft:1977xjm,Parisi:1978bj,Parisi:1978iq}, which are considered to be a pathology of perturbation theory.

 \begin{figure}\centering
     \includegraphics[scale=0.45]{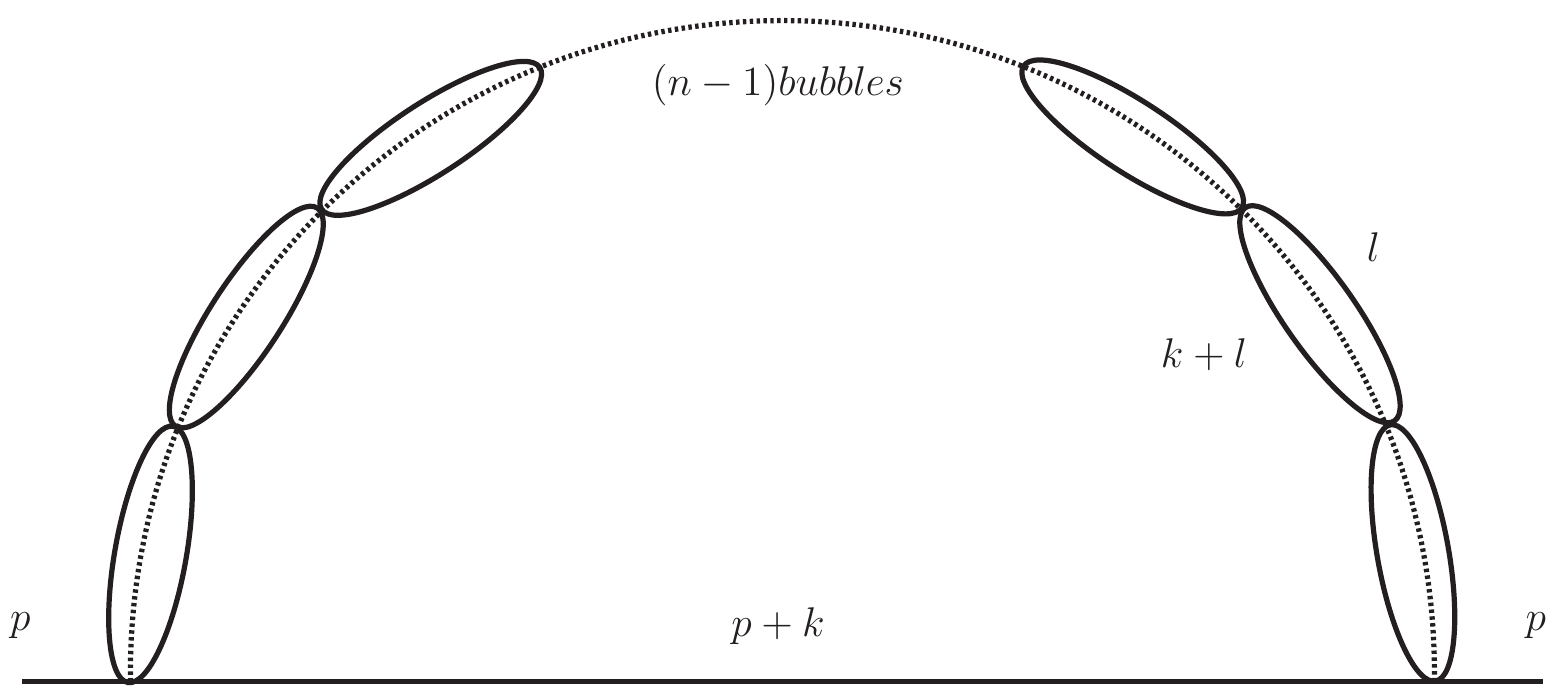}
  \caption{Skeleton diagram that gives rise to the renormalons described in~\eqref{firstfinite}-\eqref{Renormalon}. We borrow the graphics from\cite{Maiezza:2018pkk}.}
  \label{fig:Renormalon}
\end{figure}

Below, we will demonstrate how Borel-hypergeometric  resummation may cure the renormalons, but how this result incorporates into the complete non-perturbative picture of, say, low energy QCD remains an open issue. Some insights shall be mentioned below and in Sec.~\ref{Sec:end}, while now,
for the sake of completeness, we go through a brief recap of the issue.

As already sketched in Eq.~\eqref{sketch}, the divergences of series in QFT~\cite{Dyson:1952tj}  can be cured by resurgent analysis~\cite{Resurgence}. However, it may happen that even the Borel transform is divergent due to  poles on the positive real axis in which case the Laplace's transform is ambiguous. This is exactly the case with the perturbative series generated by a certain Feynman diagrams which, after the renormalization procedure, lead to Borel divergences called renormalons~\cite{tHooft:1977xjm}. Following~\cite{Maiezza:2018pkk}, consider for example the diagram in Fig.~\ref{fig:Renormalon} and denote its value at the $n$-th order of perturbation theory  as $\tilde{R}_n$
\begin{align}\label{Rn}
&\tilde{R}_n = \int \frac{d^4k}{(2\pi)^4}\frac{i}{(p+k)^2-m^2+i\epsilon}\frac{1}{(-i\lambda)^{n-1}}[B(k)]^{n}\,,\\
&B(k)\equiv \frac{(-i\lambda)^2}{2}\int\frac{d^4l}{(2\pi)^4}\frac{i}{(k+l)^2-m^2+i\epsilon}\frac{i}{l^2-m^2+i\epsilon}\,. \nonumber
\end{align}

In the large Euclidean momentum ($k$) expansion and after having absorbed the divergences in the counterterms, the first finite term, denoted as $R_n$, behaves as~\cite{tHooft:1977xjm}
\begin{equation}\label{firstfinite}
R_n\propto 1/\lambda^{n-1}\int_0^\infty \frac{[\lambda^2\beta_0/2  \lg(k^2/\mu^2)]^n}{k^6} dk^4\,,
\end{equation}
where $\lambda^2\beta_0/2  \lg(k^2/\mu^2)$ is the contribution from a single bubble of the diagram shown in  Fig.~\ref{fig:Renormalon} and $\beta_0= \frac{3}{16\pi^2}$ is the one-loop coefficient of the  $\beta-$function defined as $\beta=d\lambda/d\lg\mu$.
It is easy to show that defining a new variable $x\equiv\lg(k^2/\mu^2)$ one can rewrite Eq.~\eqref{firstfinite} as
\begin{equation}\label{evaluate}
R_n\propto 1/\lambda^{n-1}\int_0^\infty (\lambda^2\beta_0 x/2 )^n e^{-x} dx = \lambda^{n+1}\left(\frac{\beta_0}{2}\right)^{n}n! \,.
\end{equation}
Applying the Borel transform ($\mathcal{B}(\lambda^{n+1}) \equiv \frac{z^n}{n!} $) and summing over $n$,
to obtain the value for the whole infinite chain of bubbles,
one obtains
\begin{equation}
\mathcal{B}(\sum_n R_n)\propto \sum_n(\frac{\beta_0}{2}z)^n \equiv \sum_n \mathcal{B}_n(z) =\frac{1}{1-\frac{\beta_0}{2}z}\,, \label{Renormalonseries}
\end{equation}
and therefore the Borel serie diverges at
\begin{equation}\label{Renormalon}
z_{pole}=2/\beta_0\,.
\end{equation}
Since one made use of the large momentum expansion, this is called UV renormalon; similar discussion and result holds for the low momentum expansion, leading to the
Infra-Red (IR) renormalon (see~\cite{Beneke:1998ui} for a general review) that is less relevant for our discussion. For positive $\beta_0$, the pole in Eq.~\eqref{Renormalon} lies on the positive real axis, making the Laplace's transform ambiguous. As a result, there is no consistent way to perturbatively renormalize the theory when the coupling approaches the $2/\beta_0$ value. In other words, the usual QFT based on loop expansion ceases to make sense at fundamental level, even though one can still proceed in an effective way with effective operators~\cite{Parisi:1978iq} but this is not our aim, since we are interested in perturbatively  renormalizable theories.

Also, it is evident by construction that considering the  \emph{divergent part} of the integral, thus a lower power of $k$ in the denominator of Eq.~\eqref{firstfinite}, no $n!$ contributions emerge because the $ e^{-x}$ in Eq.~\eqref{evaluate} disappears. Therefore, the $\beta-$functions have nothing to do with the renormalons \textit{per se},
nevertheless one has to worry about them for finite contributions. In summary and coming back to the $\beta-$function of the $\phi^4$ model, Eq.~\eqref{Renormalon} would impose an independent cut on the coupling values in Fig.~\ref{fig:phi4} starting at $g_{pole}=2/\beta_0 = 2/3$. Curiously, this is the point where the different resummed $\beta-$functions start to deviate from each other.

A comment is now in order. While the concept of renormalon can be revisited in terms of running coupling(s), and it is a powerful method that allowed recently to make a non-trivial generalization of renormalons to theories with multiple couplings~\cite{Maiezza:2018pkk}, it is not a priori necessary at all to think in terms of RGEs to understand the divergence in Eq.~\eqref{Renormalon}. In order to estimate one bubble in Fig.~\ref{fig:Renormalon}, one deals with the $\phi^4$ ''fish-diagram'' relating it to the one-loop $\beta-$function, but this is somehow accidental in a sense that the finite part of Fig.~\ref{fig:Renormalon}, resulting from contributions of the infinite chain of bubbles, is completely independent of, say, a more accurate higher loop RGE functions describing the running. As a consequence, the notion of renormalon is independent of whether a theory has a fixed point or not and also of whether it is tuned to be at this conformal fixed point or not.

\paragraph{Can we go beyond the renormalon?} While, as we reviewed, the usual resurgence procedure sketched in Eq.~\eqref{sketch} leads  to the renormalon issue for  sufficiently large values of the coupling,
it is worth asking whether the Borel-hypergeometric resummation can solve the problem. This is exactly what we try to outline here. From Eq.~\eqref{evaluate},  let us write again the main object of the discussion here,
\begin{equation}\label{focusRen}
R_n\propto g^{n+1}\left(\frac{\beta_0}{2}\right)^{n}n!\,,
\end{equation}
where we use the coupling $g$ so that  $\beta_0=3$.
In the previous section, it was illustrated that this series is not Borel resumable due to the ambiguity at the pole $z_{pole}=2/\beta_0$. Moreover, the conventional Borel-Pad\'e resummation procedure cannot remove  the ambiguity  since
the Pad\'e approximant of the geometric series is again the geometric series and thus one does not manage to deal with the pole on the positive real axis.
This is in contrast with the instanton divergence in the Borel plane, which can be resummed  with a sufficient number of terms in the truncated series using Borel-Pad\'e resummation procedure (e.g. see Ref.~\cite{Mera:2018qte})\footnote{In the presence of the Stokes phenomenon the real axis is segmented in branch-cuts, namely different regions separated by singularities.  In each of these regions, the Borel-Pad\'e resummation has to be applied separately and often many orders in the perturbative expansion are needed to obtain a sensible result.}.
This is why when applying the Laplace transform within the Borel-Pad\'e resummation procedure, the usual ambiguity appears for renormalons but not for instantons. In any case, both the instantons and renormalons come from $n!$ divergent series and, although conceptually they are very different from each other, they represent the same kind of problem from the point of view of MG algorithm.

With this in mind,  consider the perturbative asymptotic series in Eq.~\eqref{focusRen} and keep the first few orders in $n$. Such truncated series can be resummed, exactly as done throughout this paper and the resummed   function has an asymptotic expansion precisely given by Eq.~\eqref{focusRen}. The MG approximants are quickly convergent, in full analogy with the 0-dimensional functional recalled above and studied in Ref.~\cite{Mera:2018qte}. More precisely, one obtains
\begin{equation}\label{Rnsum}
MG_{N=3,4,5}(\sum_n R_n)= -\frac{2}{\beta_0} e^{-\frac{2}{\beta_0\lambda}} \Gamma(0,-\frac{2}{\beta_0\lambda})\,,
\end{equation}
where $\Gamma(0,-\frac{2}{\beta_0\lambda})$ is the incomplete Gamma function. The expression in Eq.~\eqref{Rnsum} is exact and thus is a representation, in terms of $\Gamma-$function, of the formal and exact result emerging from the standard resurgent procedure sketched in Eq.~\eqref{sketch} (see for instance the discussion in Ref.~\cite{Resurgence}).
However, the expression in Eq.~\eqref{Rnsum} is complex, and the imaginary part is indeed known as the non-perturbative ambiguity.
This is nothing but the renormalon itself emerging from Eq.~\eqref{sketch}. Nevertheless, the fundamental observation here is that only the real part of Eq.~\eqref{Rnsum} might be meaningful as a feature of the algorithm itself. Let us clarify the meaning of this statement. Consider for example the self-interacting 0-dimensional  QFT example studied in Ref.~\cite{Mera:2018qte}. What emerges from the Borel-hypergeometric resummation is that the real part of the MG algorithm converges exactly to the exact result that is purely real, while, in addition, an imaginary part emerges as a sort of byproduct. However, it should be stressed that such an imaginary part can be consistently removed by first constructing a trans-series~\cite{ZinnJustin:2004ib} from the original series and then resummating them via MG. This combined method of trans-series with MG on the top leaves the $Re(MG)$ intact while removing the non-perturbative ambiguity and, at the end of the day, one is left with $Re(MG)$ as the only relevant result.

In the light of this argument, we speculate that the real part of Eq.~\eqref{Rnsum} is the relevant result.
A fundamental requirement in support of the above conjecture is that there must be a matching between the perturbative expansion and the full answer in the small coupling limit. When this is not the case, one is dealing with a genuine non-perturbative problem and the  MG algorithm fails to approximate the result. This situation is clearly exemplified by the 0-dimensional problem of degenerate-vacua in Ref.~\cite{Mera:2018qte}, in which the real part of the MG output fails to reproduce the exact result.
Fortunately, this does not seem to be the case for the renormalon,  which is nothing but some specific finite contributions to the pole mass (see Fig.\ref{fig:Renormalon})  and can be described perturbatively. In other words, we are arguing that the renormalon would resemble the example of the self-interacting 0-dimensional  QFT mentioned above, which indeed admits a perturbative description.

\begin{figure}
 \centering
     \includegraphics[scale=0.3]{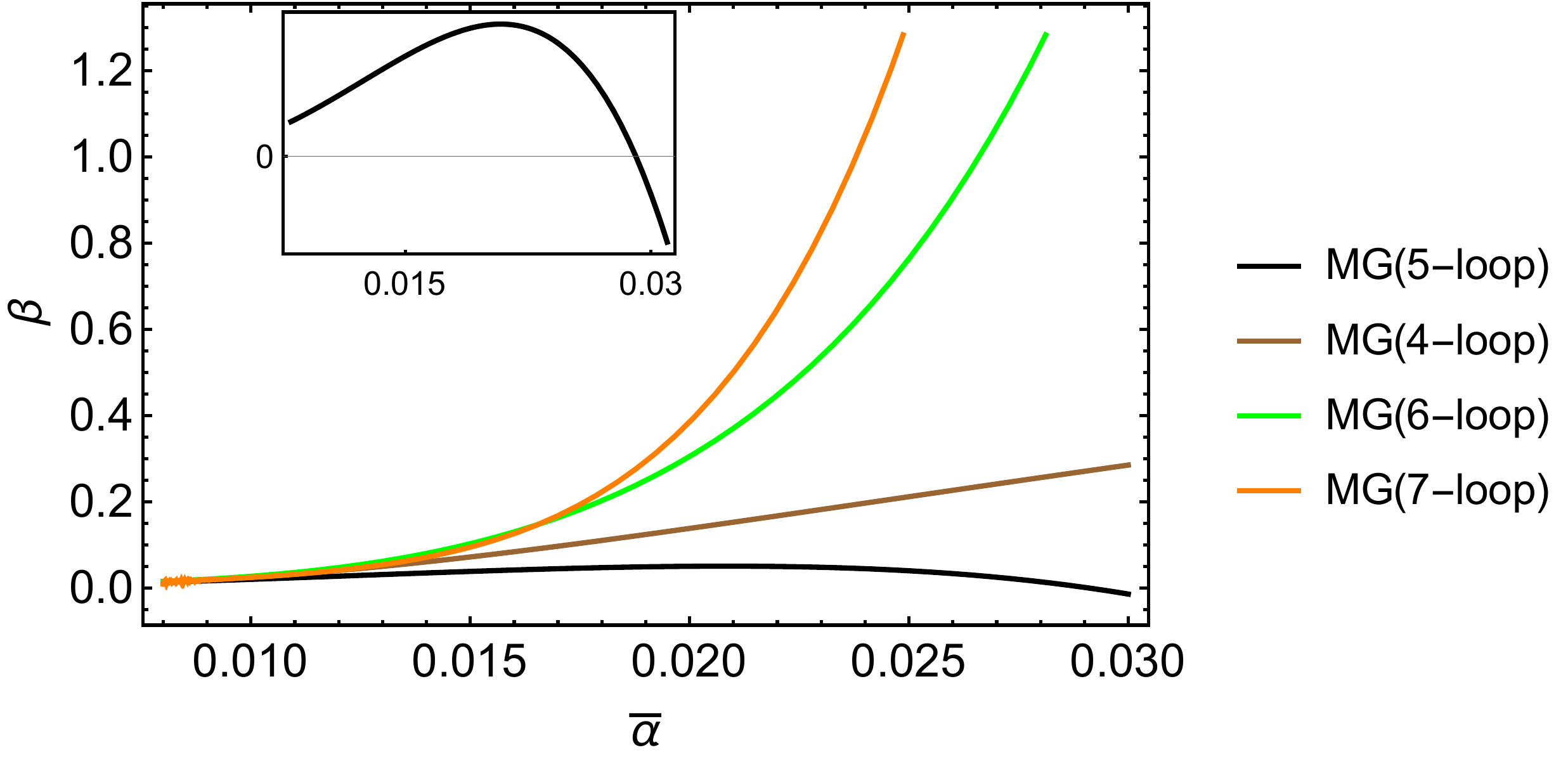}
  \caption{Meijer G-function approximants of the $\beta-$function up to 7-loop order for $U(1)$ gauge theory. Here $N_f=N_f^{crit}=160$.  The black line (zoomed) for 5-loop MG is non-convergent with the others and therefore the corresponding FP is not physical.}
  \label{U1largeN}
\end{figure}

In summary, one would need an additional tool, such as the trans-series, together with the MG algorithm in order to prove our conjecture. Still, the 0-dimensional analogy gives an insight
for considering the real part of the MG algorithm as the only meaningful object.
Finally, even though the Borel-hypergeometric resummation is a powerful tool with a  built-in self-consistency check provided by its convergence (see more details in App.~\ref{App_MeijerG}), it would be interesting to think about variations of the algorithm in such a way to obtain an independent external check. Obviously, for the 0-dimensional functional the external benchmark is the exact result, but clearly, the situation is different in a realistic QFT. Until the Borel-hypergeometricmethod is consolidated and the non-perturbative ambiguity is resolved, it is prudent to keep the renormalons as a sensitive milestone in perturbation theory. Therefore in what follows we conservatively show the renormalon pole as a benchmark.

\section{$\beta-$functions in gauge theories}\label{Sec:gauge}

We now go to the even more involved framework of gauge theories.
We shall take advantage of the whole discussion presented in the previous section for $\phi^4$ model that we directly translate here.
This is clearly possible because the mathematical construction based on Meijer-G function is generic, regardless of the underlying
physics. Moreover, the concepts of instantons and renormalons are also generic in QFT.

\subsection{$U(1)$ as a function of the number of flavors}

\begin{figure}
 \centering
     \includegraphics[scale=0.15]{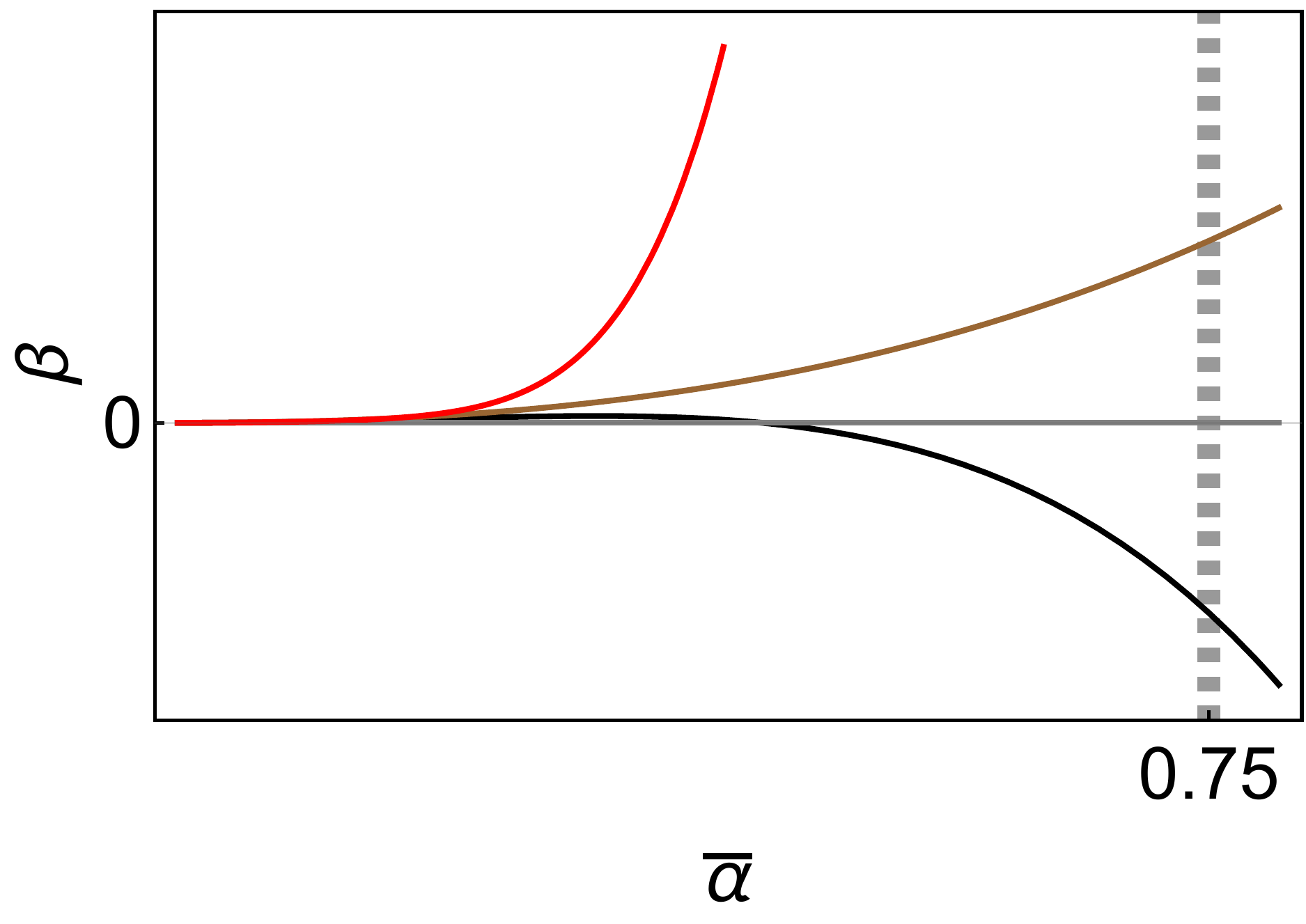}    \includegraphics[scale=0.15]{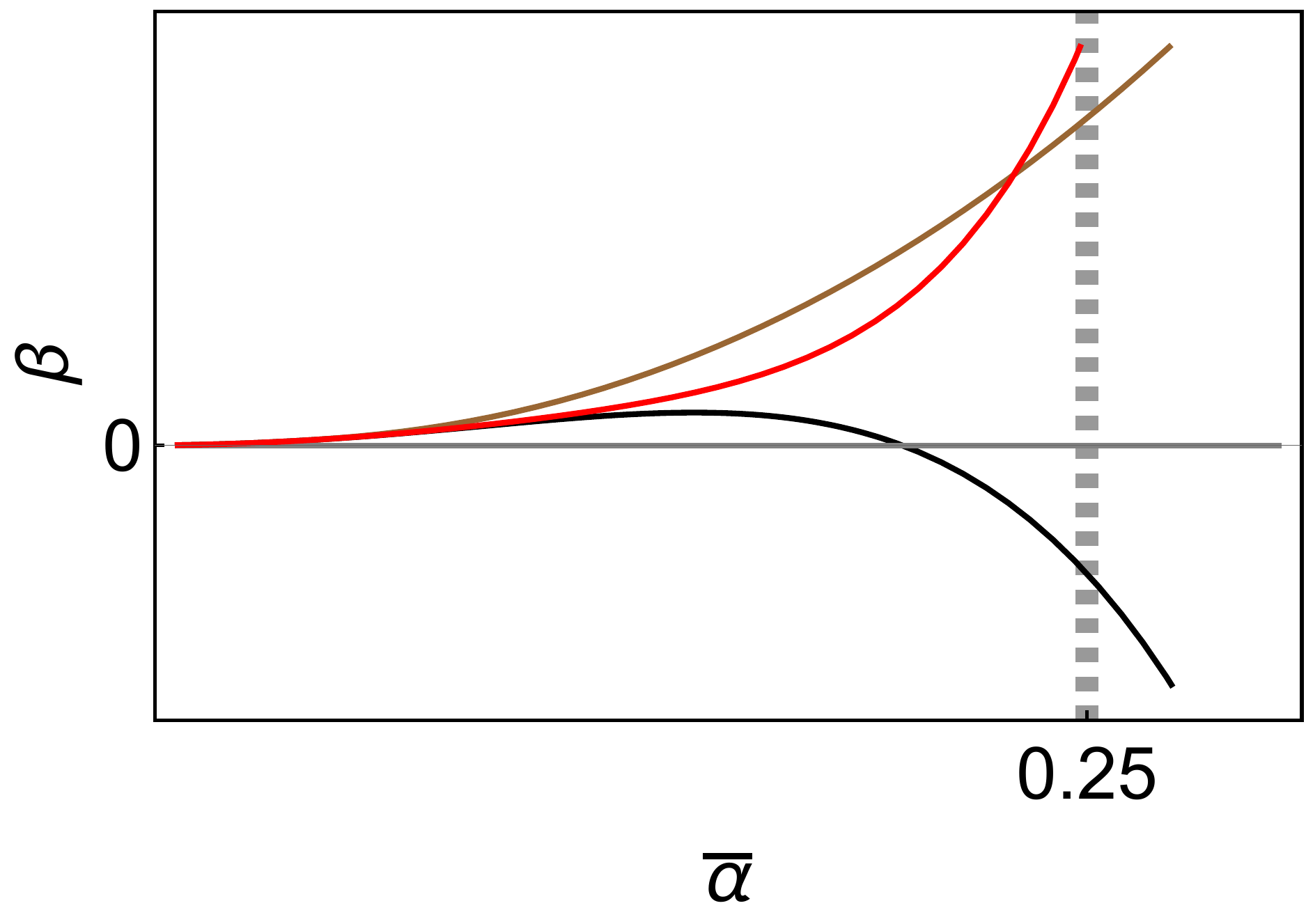}  \includegraphics[scale=0.15]{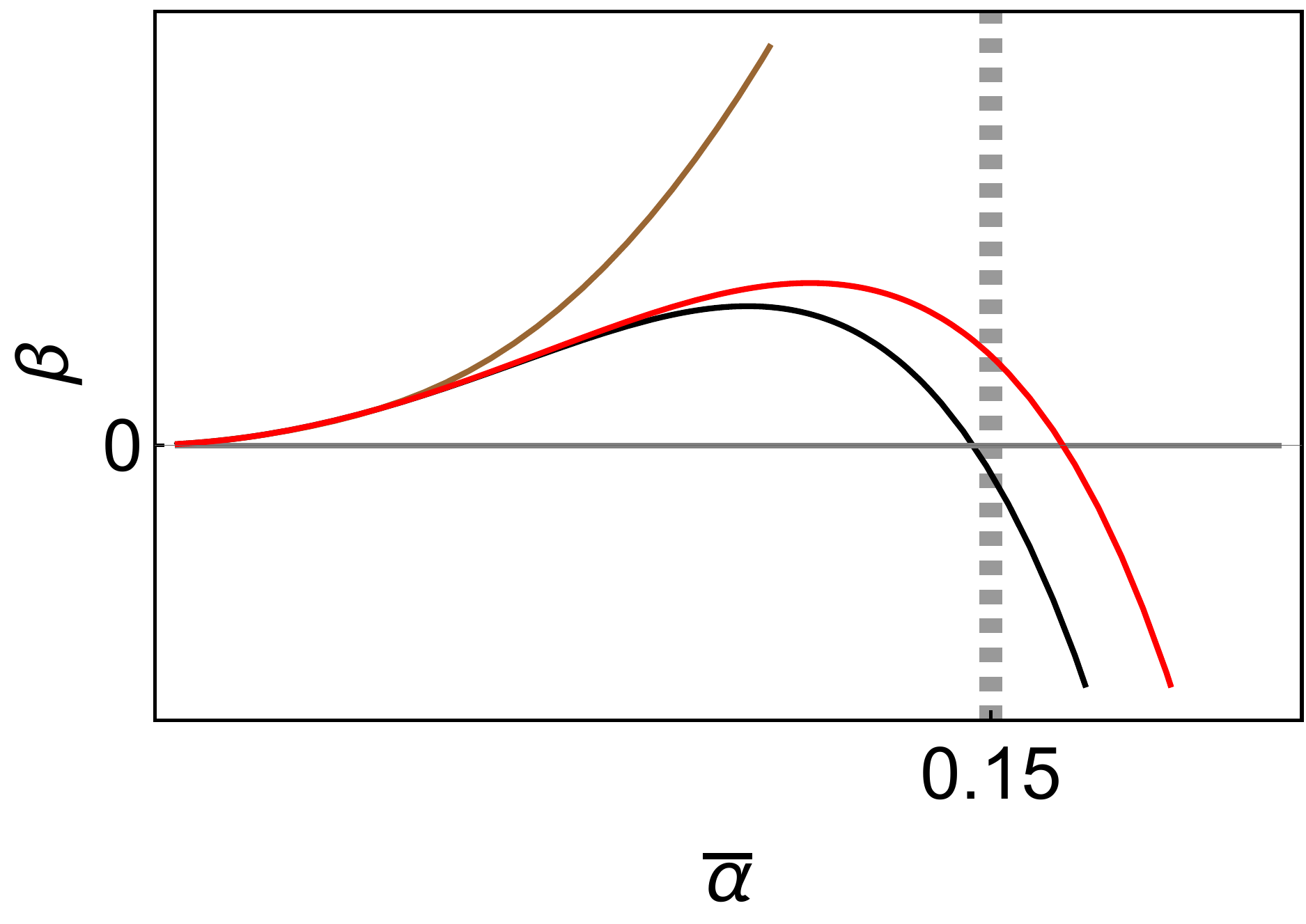}  \includegraphics[scale=0.15]{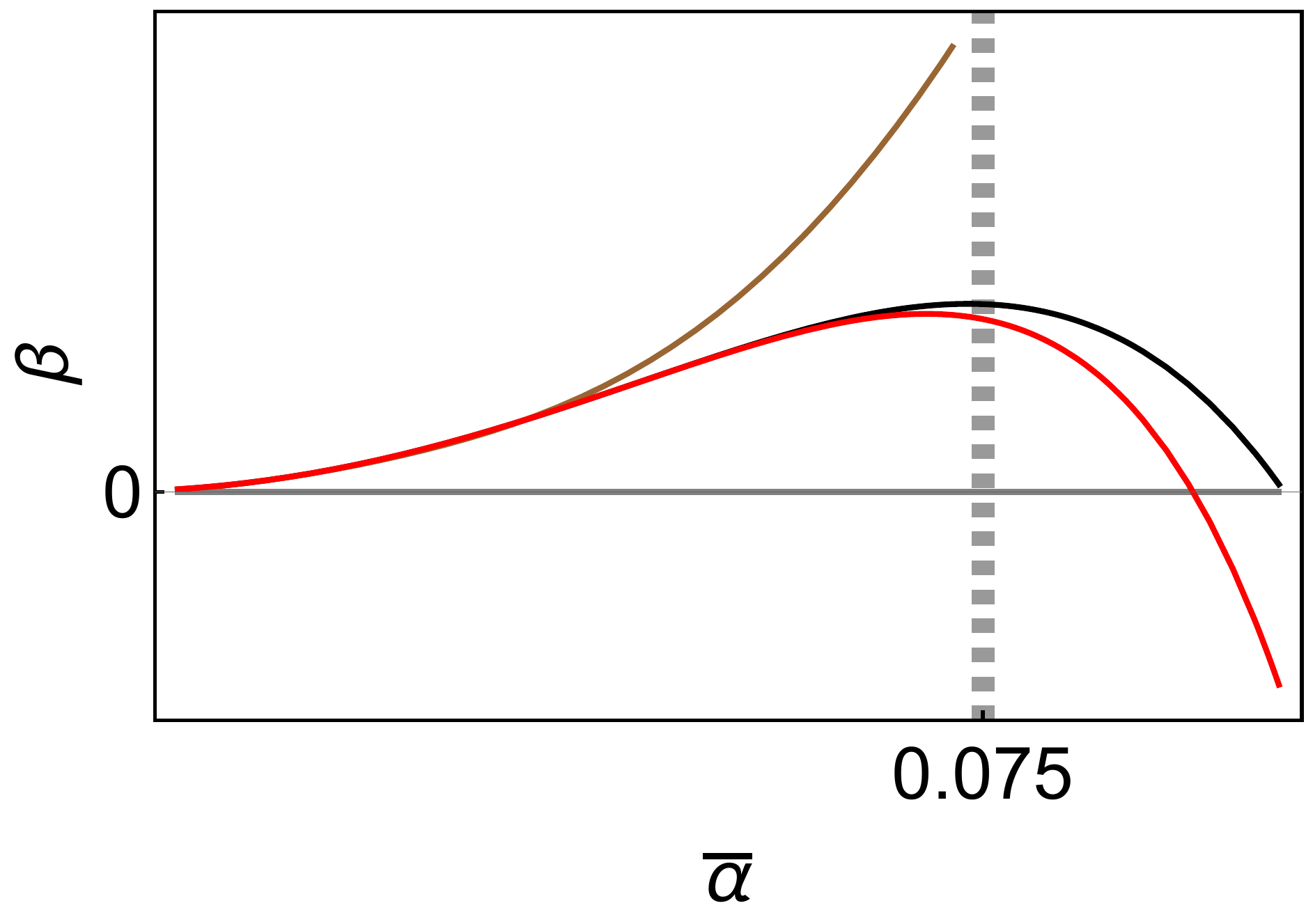}
  \caption{Low $N_f$ ($N_f=1,3,5,10$ from left to right) $\beta-$function of $U(1)$ gauge theory: perturbative (red), 4-loop MG approximant (brown) and  5-loop MG approximant (black).}
  \label{betaQEDMS}
\end{figure}

Here our aim is to analyse $U(1)$ gauge theory performing an analysis similar to the $\phi^4$ model above. The $\beta-$function of QED is known up to
5-loops and in $\overline{MS}$ scheme reads ($\bar{\alpha}=e^2/(4\pi)^2$)~\cite{Herzog:2017ohr}
\begin{eqnarray}\label{mainQED}
&{}& \beta^{U(1)}_{\overline{MS}}(\bar{\alpha}) =
\,N_f
\frac{4\ \bar{\alpha}^2}{3}
{+}
4\,  N_f \bar{\alpha}^3
-  \bar{\alpha}^4
\left[
2  \,N_f
+\frac{44}{9}  \, N_f^2
\right]
\nonumber\\
&{+}&
\left[
-46  \,N_f
+\frac{760}{27}  \, N_f^2
-\frac{832}{9}  \,\zeta_{3} \, N_f^2
-\frac{1232}{243}  \, N_f^3
\right] \bar{\alpha}^5
\nonumber\\
&{+}& \Biggl( \,N_f
\left[
\frac{4157}{6}
+128   \zeta_{3}
\right]
{+} \, N_f^2
\left[
-\frac{7462}{9}
-992   \zeta_{3}
+2720   \zeta_{5}
\right]
\nonumber\\
&{+}& \, N_f^3
\left[
-\frac{21758}{81}
+\frac{16000}{27}   \zeta_{3}
-\frac{416}{3}   \zeta_{4}
-\frac{1280}{3}   \zeta_{5}
\right]
{+} \, N_f^4
\left[
\frac{856}{243}
+\frac{128}{27}   \zeta_{3}
\right]
\Biggr)\bar{\alpha}^6   \,, \nonumber \\
\end{eqnarray}
where $N_f$ is the number of Dirac fermions.
In order to reach the 7-loop precision in analogy with the analysis done for the $\phi^4$ model, we take advantage of the large $N_f$ expansion. We estimate the 6,7-loop contributions
to the $U(1)$ $\beta-$function by expanding the exact all-order leading $1/N_f$ result~\cite{Holdom:2010qs} so that, for a large enough $N_f$, the 6,7-loop terms will be approximated as
\begin{align}\label{sixsevenQED}
\Delta\beta_{6,7}= &\bar{\alpha}^7 N_f^5 \left(-\frac{11264 \zeta_3}{1215}+\frac{512 \pi ^4}{6075}+\frac{16064}{3645}\right)+ \nonumber \\
&\bar{\alpha}^8 N_f^6 \left(-\frac{78848 \zeta_3}{6551}+\frac{4096 \zeta_5}{243}+\frac{4288}{729}-\frac{5632 \pi ^4}{32805}\right)\,.
\end{align}

With Eq.~\eqref{mainQED} and the extra terms in Eq.~\eqref{sixsevenQED}, we build the MG approximants on the same level as in $\phi^4$ model. Of course, these approximants will be trustable only above some critical $N_f^{crit}$ but such $N_f^{crit}$ certainly exists since the known parts of the  6 and 7 loop terms above numerically are $\mathcal{O}(1)$ and do not have fine-tuned cancellations. Therefore, the larger $N_f$ becomes, these terms will approximate the full 6 and 7 loop coefficients more and more accurately. We estimate $N_f^{crit}\approx160$, by requiring that the terms with the highest power of $N_f$ in 4- and 5-loop coefficients in Eq.~\eqref{mainQED} dominate over the terms with the lower power of $N_f$ and assuming that this domination is also sufficient for 6- and 7-loop coefficients.

Our result is shown in Fig.~\ref{U1largeN}, where we have selected from the table of the MG approximants, the ones that maximize the convergence. In particular, as we increase the loop order, there is a progressive convergence among the MG approximants of 4,6,7-loops. We should stress that the convergence is not as good as in $\phi^4$ model shown in Fig.~\ref{fig:phi4}, although it is sufficient to establish the fact that the 5-loop MG approximant is anomalous in the sense that it is not convergent with the others. Therefore, the corresponding FP predicted by it (the black line in Fig.~\ref{U1largeN}) is not reliable and we conclude that the $\beta-$function of $U(1)$ gauge theory does not have an FP for $N_f^{crit}$. We have also checked that for the larger number of flavors $N_f>N_f^{crit}$  the FP does not develop.

\begin{figure}
 \centering
    \includegraphics[scale=0.4]{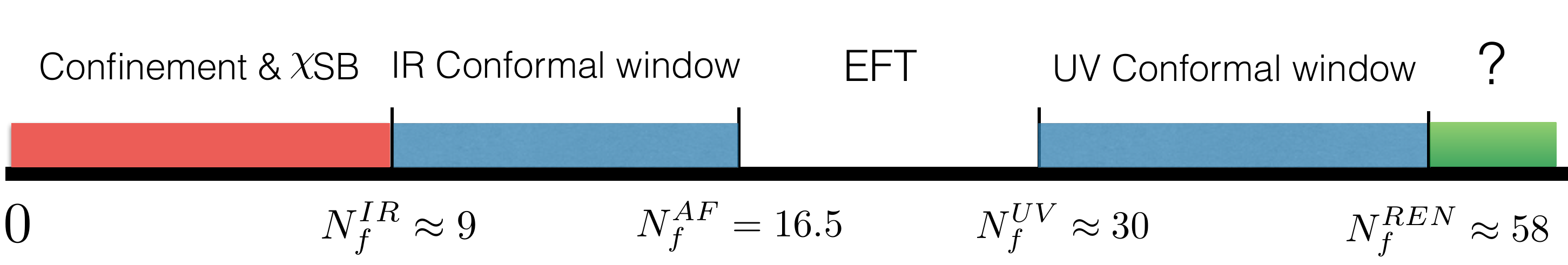}
  \caption{ $SU(3)$ gauge theory behavior as a function of $N_f$.}
  \label{phaseQCD}
\end{figure}

Following the logic in Sec.~\ref{Sec:phi}, one has to also consider the (UV) renormalon (see especially the discussion in Subsection~\ref{Subsec:ren})
and, with the normalization of $\beta-$function in Eq.~\eqref{mainQED},
one has\footnote{The UV renormalon is written as either $1/\beta_0$ or $2/\beta_0$ depending on the definition of $\beta-$function, respectively $\beta=dg/d\lg\mu^2$ or $\beta=dg/d\lg\mu$. In Subsec.~\ref{Subsec:ren} we used the latter, while in Eq.~\eqref{mainQED} we use the former.}
\begin{equation}\label{largeNQED_Ren}
R_{N_f^{crit}}^{UV}=\frac{1}{\beta_0}=\frac{3}{4 N_f^{crit}}\approx 0.005\,.
\end{equation}
From this equation it is clear that
the first UV renormalon is approaching the origin as we increase $N_f$ and in the strict $N_f\to \infty$ limit it occurs at zero coupling. For this reason, the $1/N_f$ expansion has to be applied with care as $N_f$ has to be large enough for the expansion to be valid and yet not too large so that the first UV renormalon does not occur too close to the origin.
Also, it is interesting to notice that for $U(1)$ gauge theory the topology of the diagrams leading to the renormalon is exactly the same as the one that provides the leading $1/N_f$ term of the $\beta-$function in the large $N_f$ expansion~\cite{Antipin:2018zdg}. The difference, as usual, is that while the renormalons come from the finite part of the diagram, as we discussed in detail in Subsec.~\ref{Subsec:ren}, the $\beta-$function comes from the logarithmically divergent part.

As $N_f$ decreases and in particular for $N_f\ll N_f^{crit}$, the $1/N_f$ expansion is not valid anymore so that the 6- and 7- loop terms in Eq.~\eqref{sixsevenQED} are not reliable. We therefore can build the MG approximants only up to 5-loops and we show our result in Fig.~\ref{betaQEDMS} for $N_f=1,3,5,10$. However, some useful information can be still extrapolated from the large $N_f$ limit illustrated in Fig.~\ref{U1largeN}. We notice that the behavior exhibited at large $N_f$ by both 4 and 5-loop MG approximants is the same even for low $N_f$. Since we have shown that the 5-loop MG approximant is clearly a sham, one may extrapolate large $N_f$ conclusions to a smaller number of flavors. Therefore, from Fig.~\ref{betaQEDMS} we conclude that no fixed points emerge in $U(1)$ gauge theory from the present analysis. Finally, by reevaluating Eq.~\eqref{largeNQED_Ren} for $N_f=1,3,5,10$, the position of respective UV renormalons is marked in Fig.~\ref{betaQEDMS} with the dashed grey vertical line.

\subsection{$SU(3)$ as a function of the number of flavors}

We now turn our attention to the non-Abelian SU$(N)$ gauge theories and analyse the exemplary $SU(3)$ theory in our numerics. The 5-loop $\beta$-function for the $SU(3)$ gauge theory reads ($a=g_{gauge}^2/(4\pi)^2$)~\cite{Herzog:2017ohr}
\small{
\begin{align}
&\beta^{SU(3)}_{\overline{MS}}(a)= -a^2 \left(11-\frac{2 N_f}{3}\right)-a^3
   \left(102-\frac{38 N_f}{3}\right) -a^4
   \left(\frac{325 N_f^2}{54}-\frac{5033 N_f}{18}+\frac{2857}{2}\right)  - \nonumber \\
   &  a^5 \left[\frac{1093
   N_f^3}{729}+N_f^2 \left(\frac{6472 \zeta_3}{81}+\frac{50065}{162}\right)+N_f \left(-\frac{6508 \zeta_3}{27}-\frac{1078361}{162}\right)+3564\zeta_3+\frac{149753}{6}\right] - \nonumber \\
   &a^6 \Bigg[N_f^4 \left(\frac{1205}{2916}-\frac{152 \zeta_3}{81}\right)+N_f^3 \left(-\frac{48722\zeta_3}{243}+\frac{460 \zeta_5}{9}-\frac{630559}{5832}+\frac{809 \pi ^4}{1215}\right)+ \nonumber \\
   & N_f^2
   \left(\frac{698531\zeta_3}{81}-\frac{381760 \zeta_5}{81}+\frac{25960913}{1944}-\frac{5263 \pi ^4}{405}\right)+\nonumber \\
   & N_f
   \left(-\frac{4811164\zeta_3}{81}+\frac{1358995 \zeta_5}{27}-\frac{336460813}{1944}+\frac{6787 \pi ^4}{108}\right)-\nonumber \\
   &   288090 \zeta_5+\frac{621885\zeta_3}{2}-\frac{9801 \pi
   ^4}{20}+\frac{8157455}{16}\Bigg]\,.
\end{align}}
The phase diagram  of this theory as a function of $N_f$ that  emerges from our subsequent analysis is shown in Fig.~\ref{phaseQCD}.
As the middle reference point, we use the value of $N_f$ where asymptotic freedom is lost i.e. where the one-loop coefficient of the beta function $\beta_0=0$. This value is $N_f^{AF}=\tfrac{11 N}{2}$ and for the $SU(3)$ gauge group is equal to 16.5. Decreasing $N_f$ slightly below this value, one achieves the perturbative Banks-Zaks infrared fixed point (IRFP)~\cite{Banks:1981nn} , which to the two-loop level is simply given by $a^*=-\beta_0/\beta_1$ and is guaranteed to be perturbative by tuning $N_f$ such that $|\beta_0|\approx 0$. Also, since $\beta_0$ and $\beta_1$ have different signs, the FP value is physical $a^*>0$ \footnote{ For the most recent study of this perturbative IRFP based on the state-of-the-art 5-loop beta function see~\cite{Ryttov:2016ner}.}. As we lower $N_f$ further, this IRFP becomes more and more strongly coupled and at some value $N_f^{IR}$ disappears so that at the lower energy the theory is expected to confine and break chiral symmetry.

\begin{figure}
 \centering
     \includegraphics[scale=0.15]{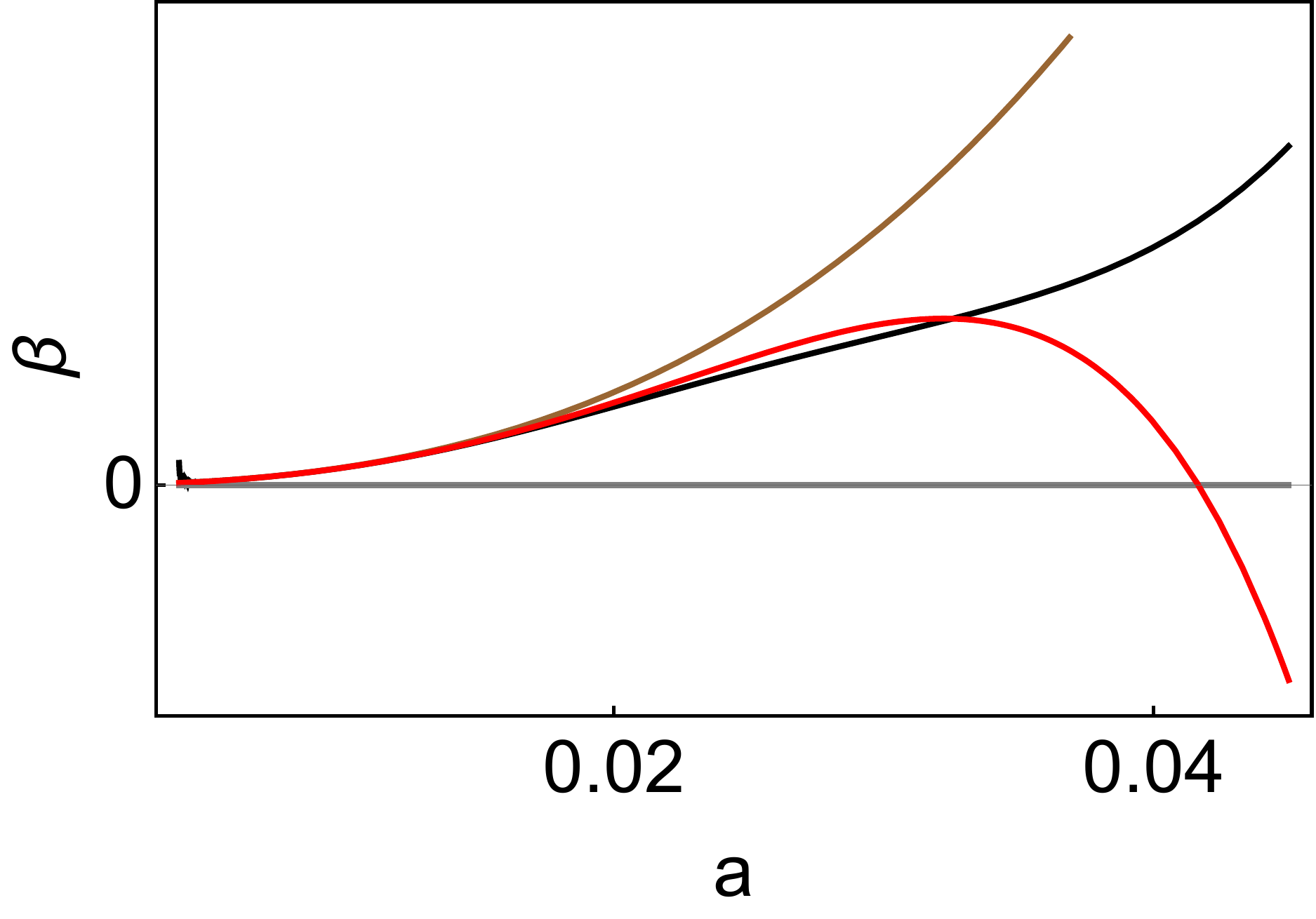}    \includegraphics[scale=0.15]{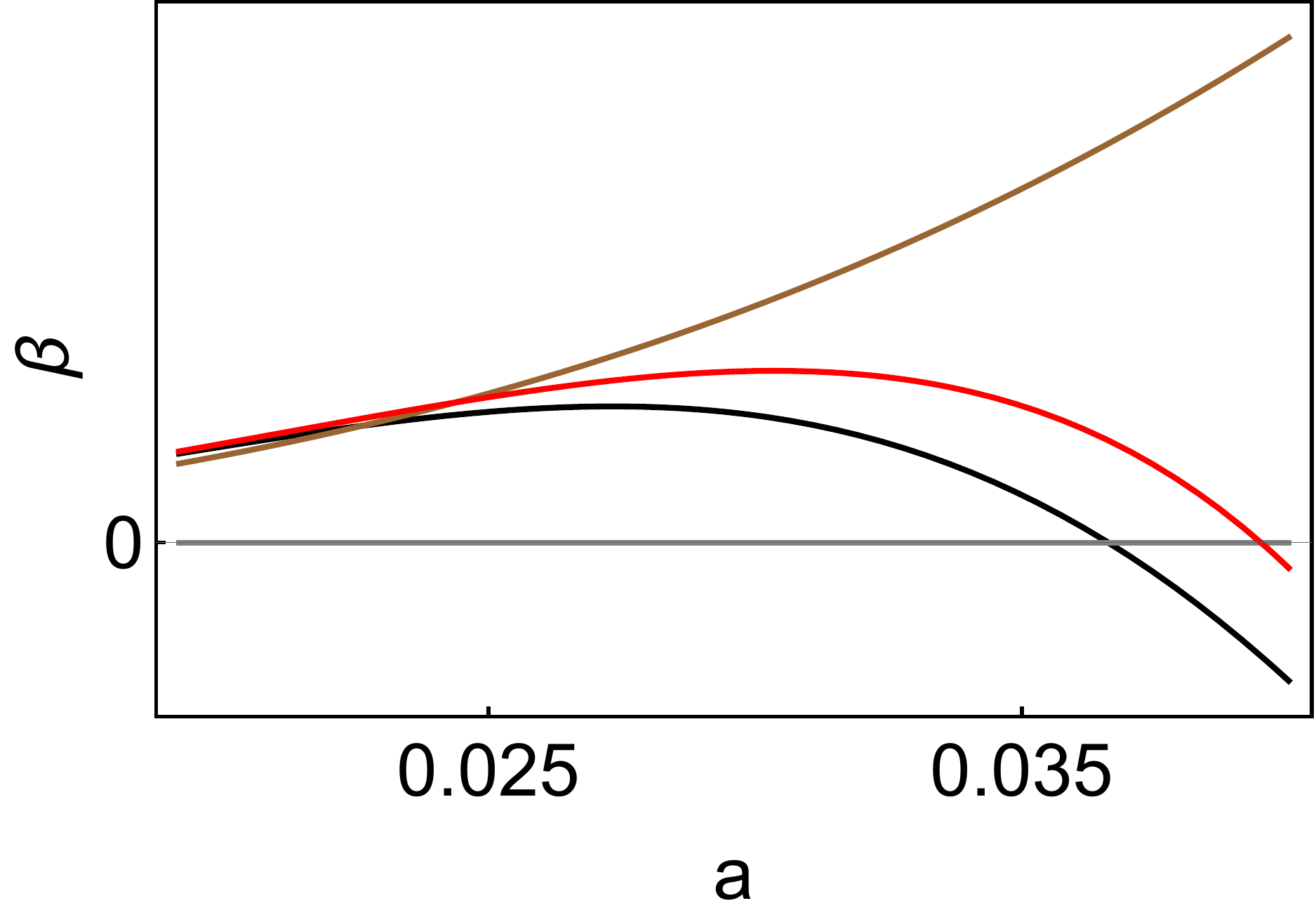}  \includegraphics[scale=0.15]{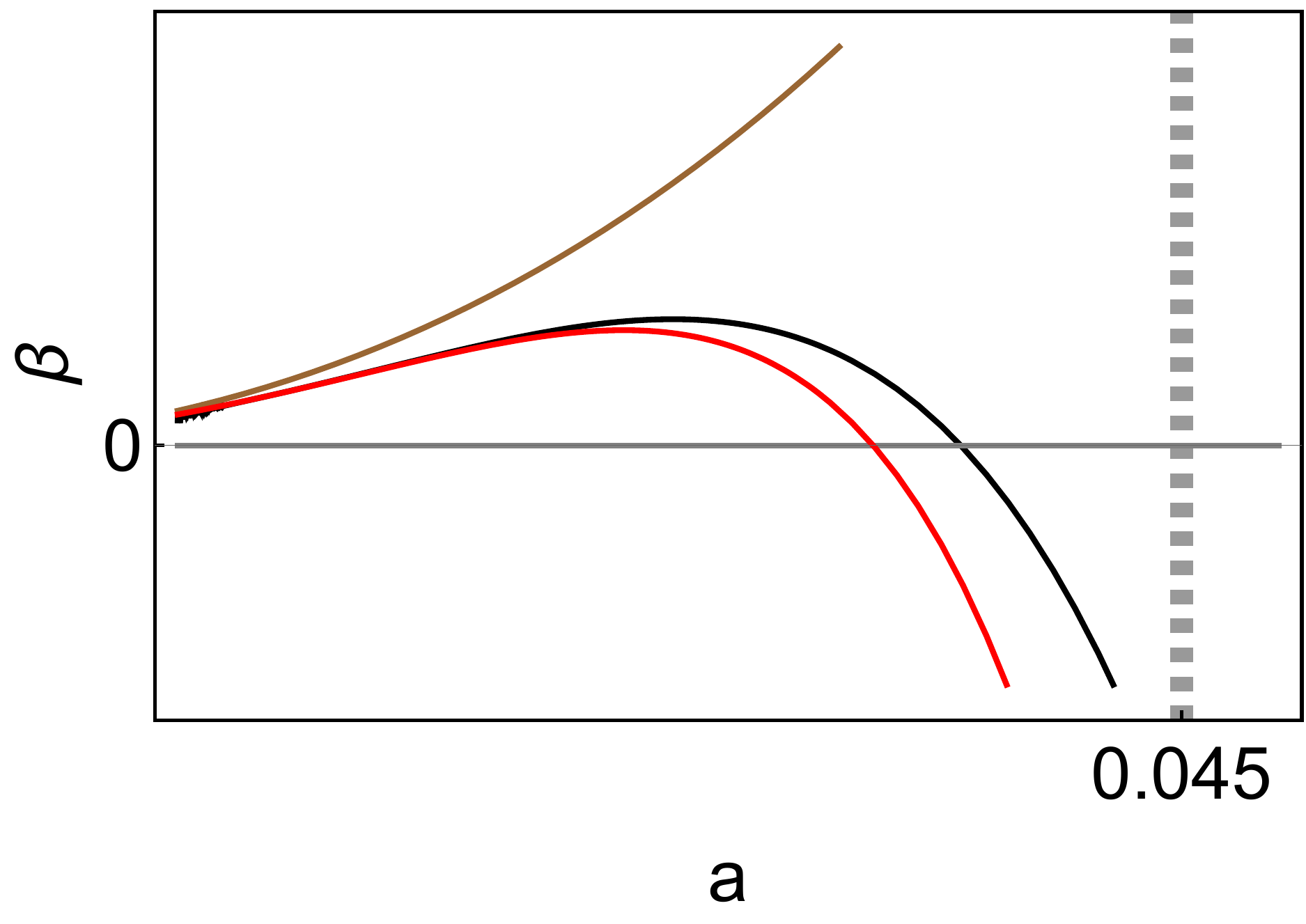}  \includegraphics[scale=0.15]{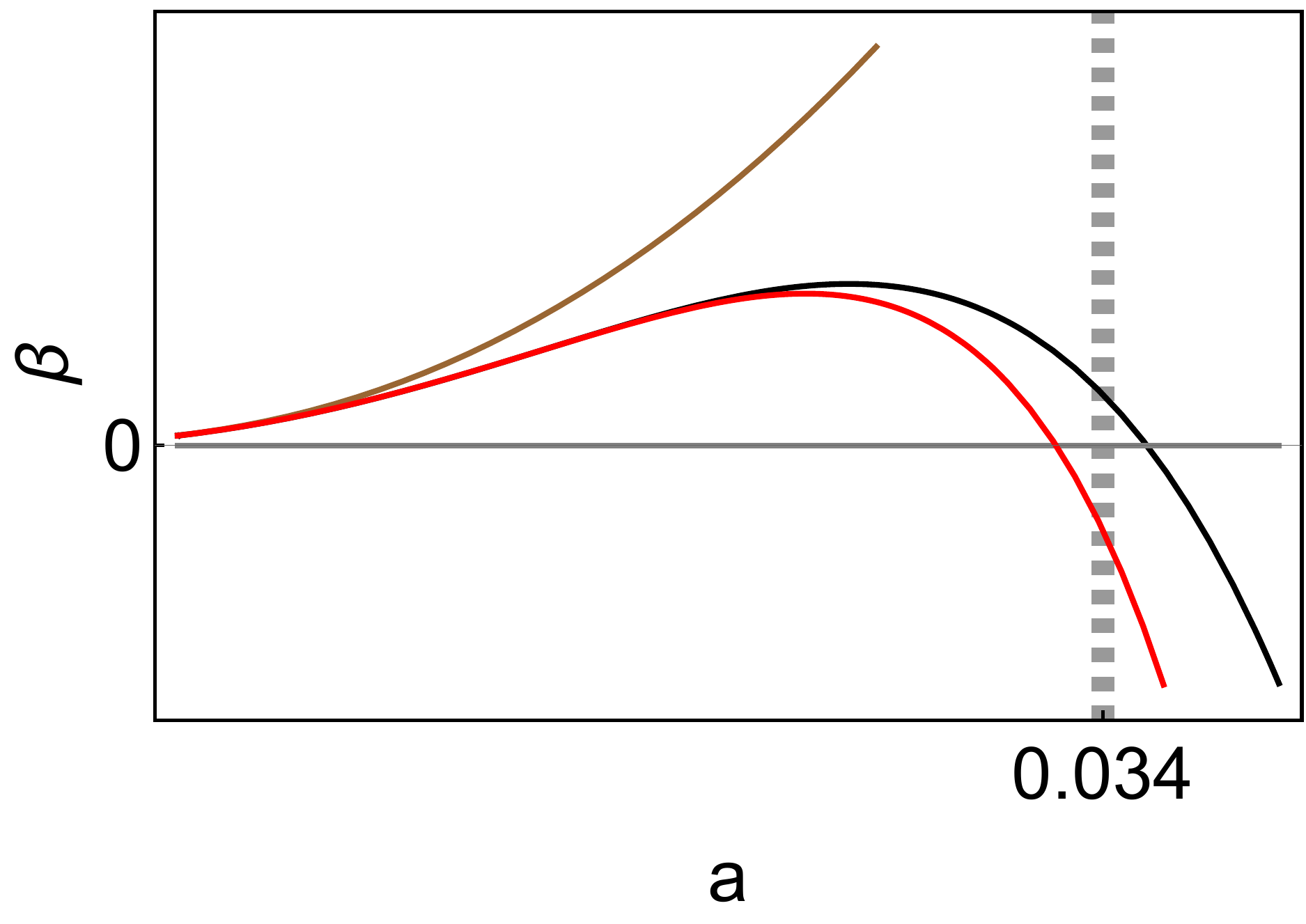}
  \caption{The $\beta-$function of $SU(3)$ gauge theory for $N_f=25,30,50,60$ (from left to right): perturbative (red), 4-loop MG approximant (brown) and  5-loop MG approximant (black).}
  \label{betaQCDMShigh}
\end{figure}

Going towards the larger values of $N_f$ from the reference $N_f^{AF}$ value, the $\beta_0$ changes sign and the potential non-trivial fixed point will be ultraviolet one. However, it was demonstrated long ago~\cite{Caswell:1974gg} that no ultraviolet FP emerges just above $N_f^{AF}$ and so, by continuity, there will be a segment in $N_f$ where the theory will be in "non-abelian QED" phase with the Landau pole at high energies and free theory at low energy. Since the UV completion of such theory is unknown, the low energy theory can be viewed as an effective field theory (labeled in Fig.~\ref{phaseQCD} as ``EFT'') featuring free Gaussian infrared fixed point. Then we may expect that there is a critical value $N_f^{UV}$ above which the non-trivial UVFP might develop.

So, now our goal is to find the $N_f^{IR}$ and $N_f^{UV}$ values  from the MG algorithm using the Meijer G-functions. We start with $N_f^{UV}$ and, analogously to the analysis of the $U(1)$ gauge theory, we may resort to the large $N_f$ limit~\cite{Gracey:1996he} and from the 4- and 5-loop coefficients to estimate the value of $N_f^{crit}$ for which the leading-$N_f$ terms dominate. We obtain $N_f^{crit}\simeq 900$. Above this value, the large-$N_f$ expansion is valid and we may use the known leading 6- and 7-loop terms
which are
\begin{align}\label{sixsevenQCD}
\Delta\beta_{6,7}=&a^7 N_f^5
   \left(-\frac{1040 \zeta_3}{729}-\frac{2069}{10935}+\frac{304 \pi ^4}{18225}\right)+\nonumber \\
   &a^8 N_f^6 \left(-\frac{8744 \zeta_3}{19683}+\frac{1216 \zeta_5}{729}-\frac{349}{4374}-\frac{260 \pi ^4}{19683}\right) \ .
\end{align}

\begin{figure}
 \centering
     \includegraphics[scale=0.3]{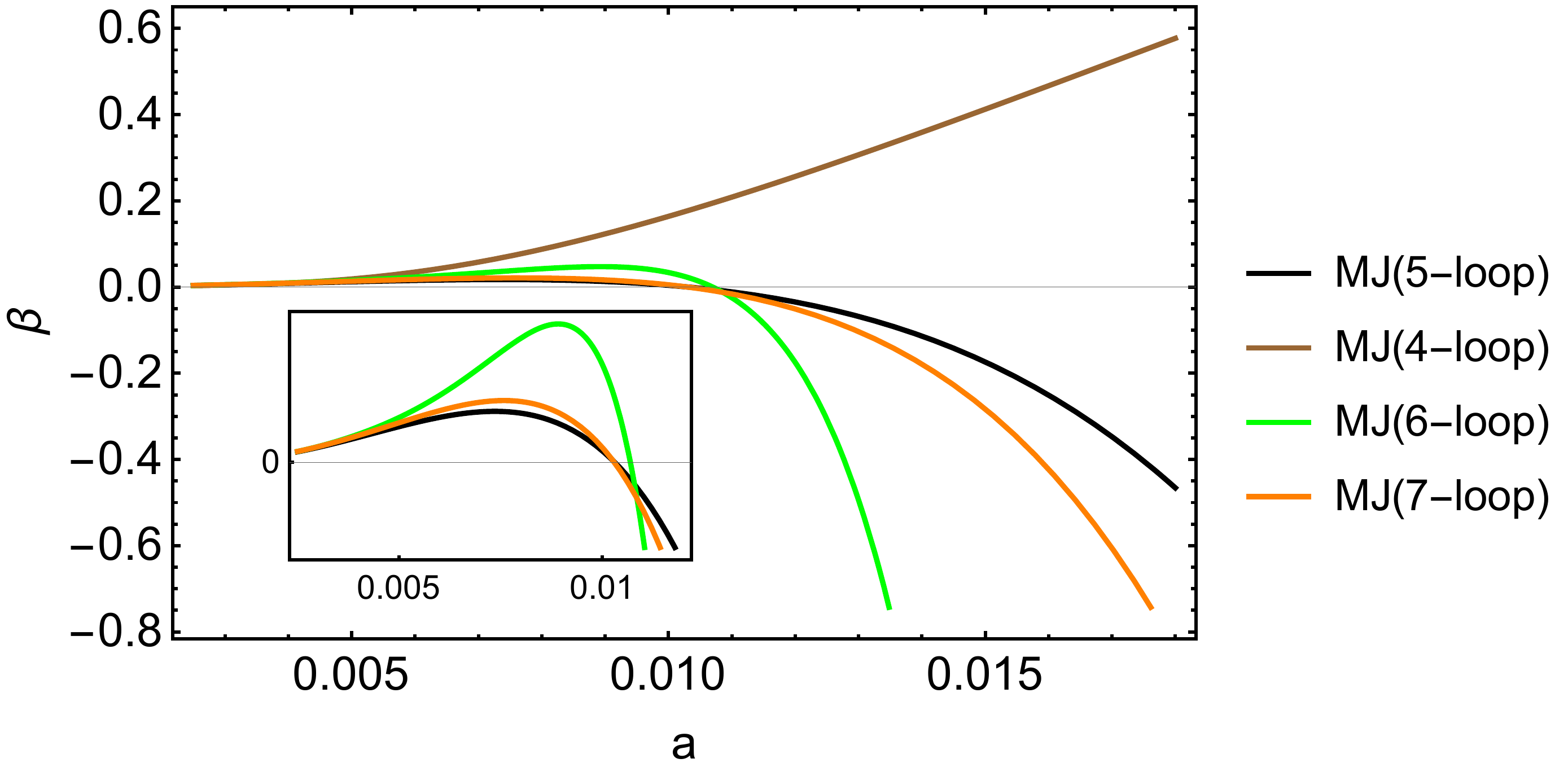}
  \caption{Meijer G-function approximants of the $\beta-$function up to 7-loop order for $SU(3)$ gauge theory. Here $N_f=N_f^{crit}=900$.  The brown 4-loop MG is non-convergent with the higher loop MG approximants which predict the UVFP.}
  \label{QCDlargeN}
\end{figure}

In Fig.~\ref{QCDlargeN}, we show the resummed  $\beta$-functions at 4- and 5-loops, together with the estimate of the 6- and 7-loops contribution in the large $N_f$ limit.  We see that unlike the QED case shown in Fig.~\ref{U1largeN}, there is a UV fixed point first appearing in the 5-loop approximation and consistently confirmed at  6 and 7 loops. Notice that the 4-loops Meijer G-function approximant  cannot be trusted in this case, since its behavior is not converging at all to 5,6 and 7 loop estimates.

A remarkable thing is that the 5-loop MG $\beta-$function provides a meaningful result, and we argue that this approximant can be extrapolated to lower $N_f$ where a 5-loop approximation is the best that one can do. In Fig.~\ref{betaQCDMShigh} we show the MG $\beta$-functions for 4 (brown) and 5 (black) loops for $N_f=25,30,50,60$. The dashed grey vertical line for $N_f=50$ and 60 represents the position of the renormalon pole and beyond this point, standard perturbative quantization might not be adequate.
Although we have shown in Subsection~\ref{Subsec:ren} that MG itself may resum the renormalon series, the relation of such summation with other non-perturbative effects is still not clear. Staying on the conservative side, we, therefore, learn that starting from $N_f\approx 58$ the MG prediction should be interpreted with care, since additional non-perturbative methods may be needed to cast light on the ultimate ultraviolet completion of these theories -- e.g. lattice quantization. In Fig.~\ref{phaseQCD}, the start of this region is colored in green and assigned the "?" sign.  Instead, for $N_f=25$ and 30 the renormalon pole is far to the right beyond the corresponding plots. For $N_f\approx 30$ the ultraviolet fixed point appears for the last time and disappears as we lower $N_f$ further. Interestingly, using a different method, this value of $N_f$ was estimated in \cite{Holdom:2010qs} as the value where the $1/N_f$ expansion starts to be reliable and the UVFP is predicted. This confirms and justifies our extrapolation procedure from the large $N_f$ to lower $N_f$ based just on the 5-loop MG result. In summary, we estimate the extent of the "UV conformal window" \cite{Antipin:2017ebo} at least as the line segment $N_f\approx(30,58)$.

\begin{figure}
 \centering
     \includegraphics[scale=0.16]{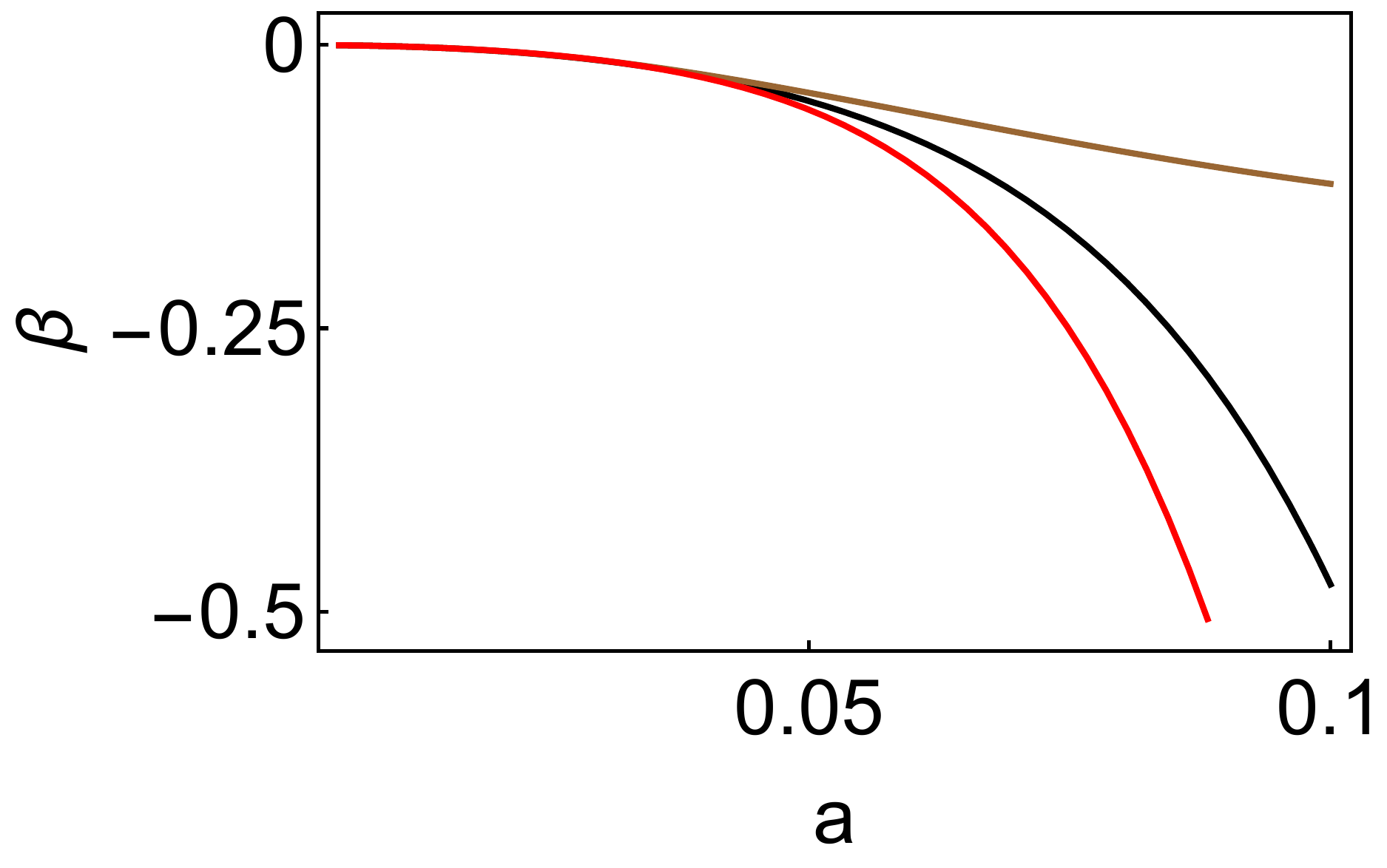}    \includegraphics[scale=0.16]{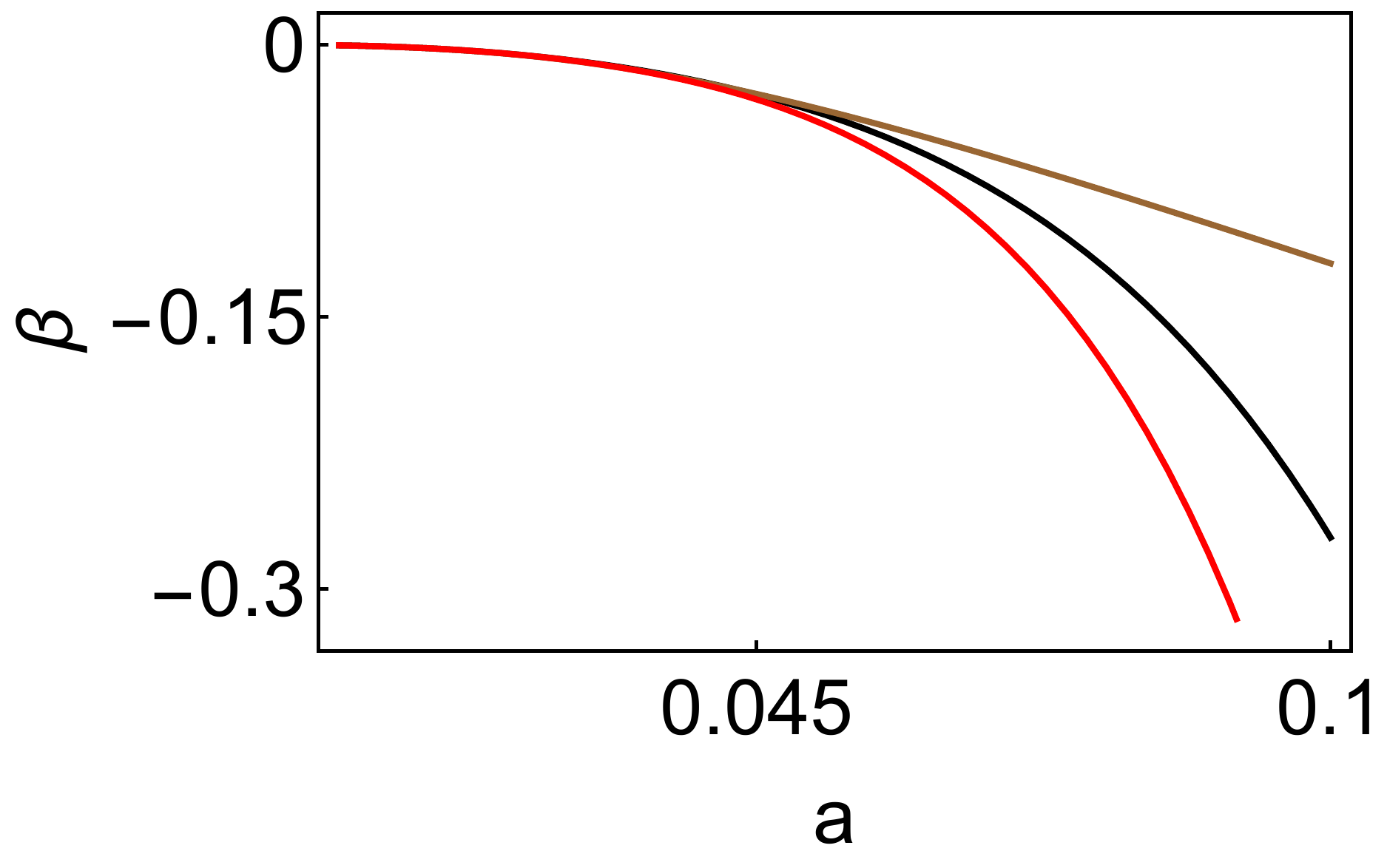}  \includegraphics[scale=0.14]{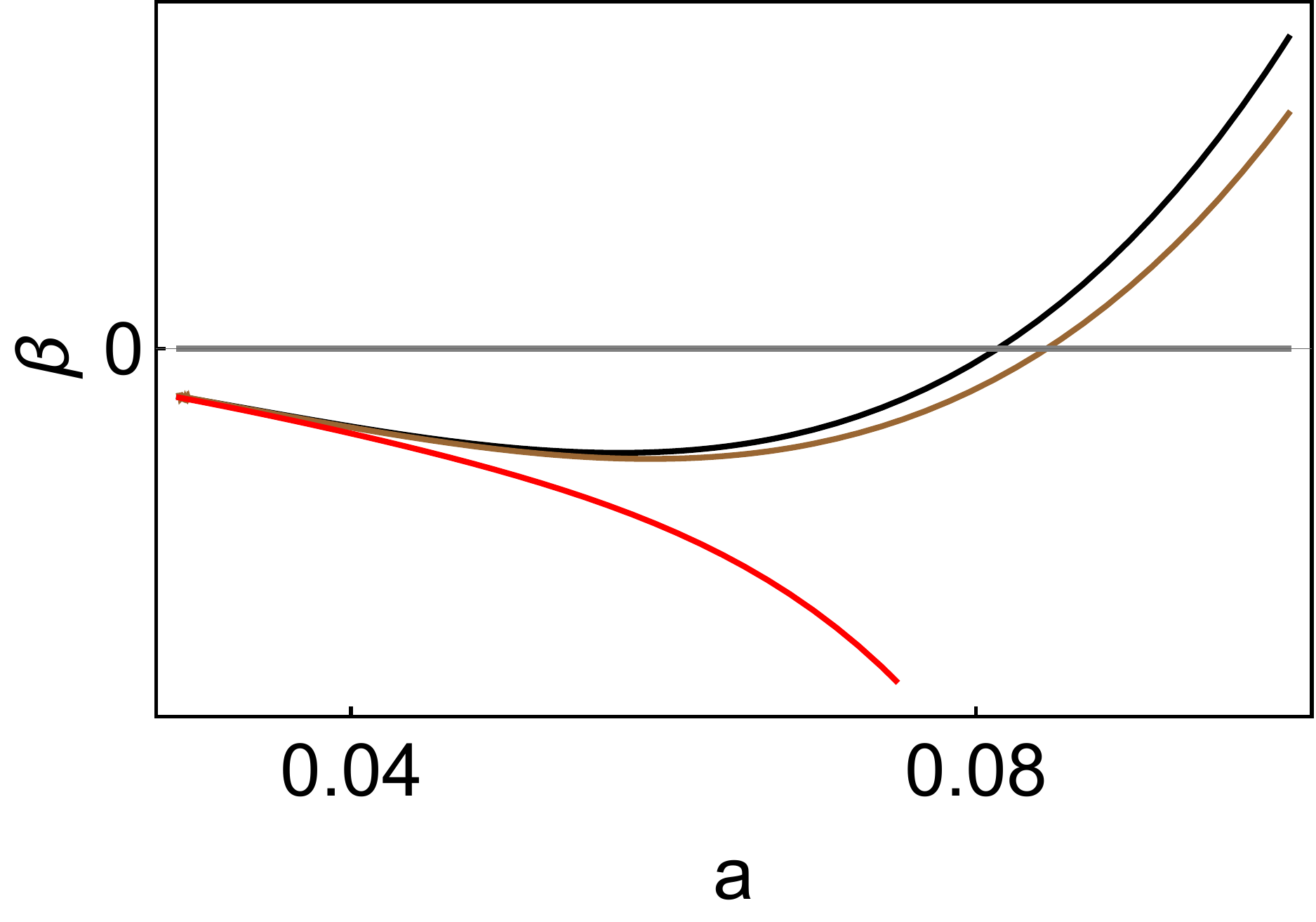}  \includegraphics[scale=0.14]{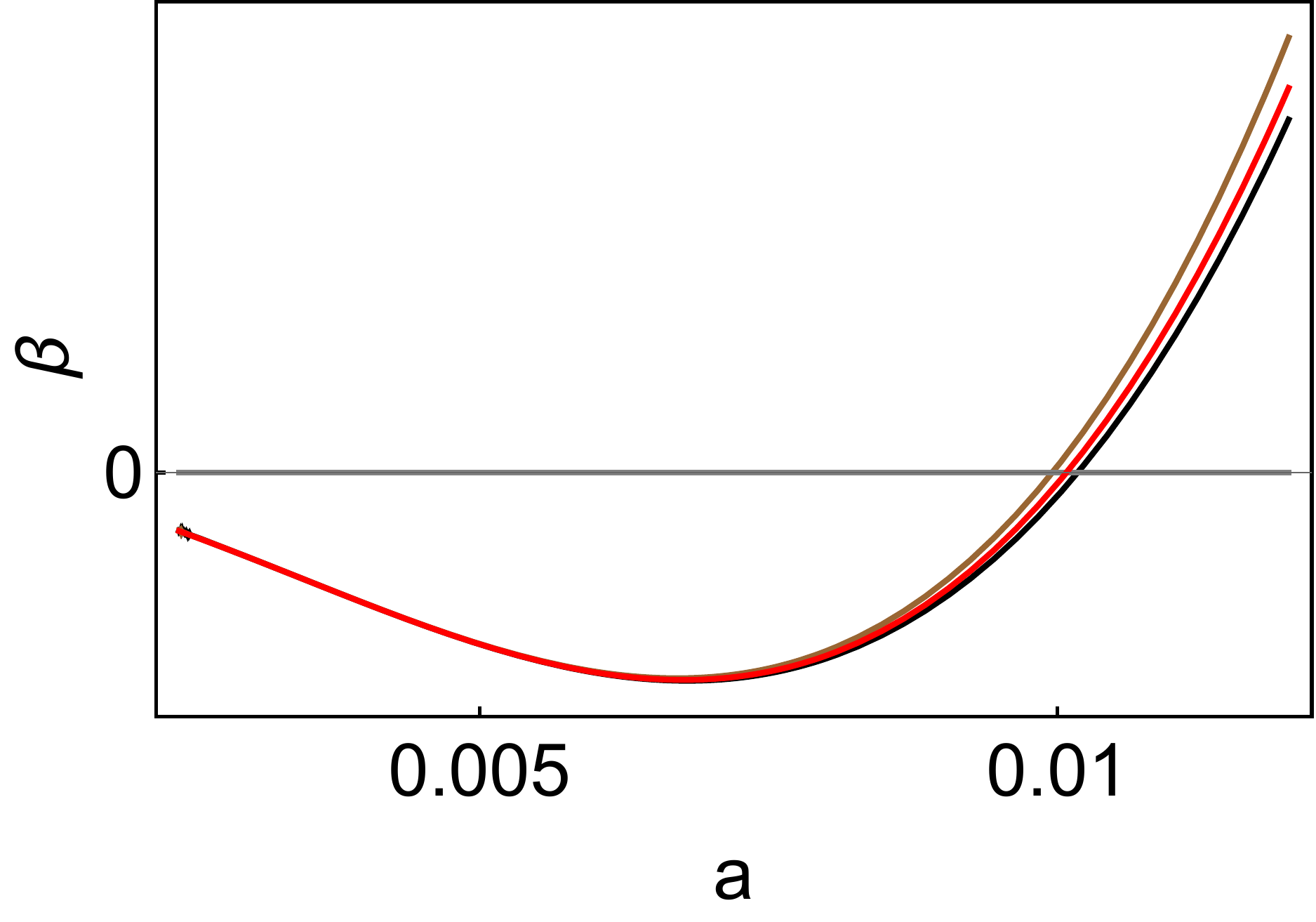}
  \caption{Low $N_f$ ($N_f=1,3,9,15$ from left to right) $\beta-$function of $SU(3)$ gauge theory: perturbative (red), 4-loop MG approximant (brown) and  5-loop MG approximant (black).}
  \label{betaQCDMS}
\end{figure}

For low values of $N_f$, in Fig.~\ref{betaQCDMS} we show the behavior of the beta functions for $N_f=1,3,9,15$ using the 4- and 5-loop MG result. Remarkably for $N_f<9$ we found that there is no infrared fixed point which fixes our final unknown on the phase diagram in Fig.~\ref{phaseQCD}, $N_f^{IR}\approx 9$. Below this value, in  $SU(3)$ gauge theory, a phase transition and spontaneous breaking of the chiral symmetry by the vacuum condensates is expected.
Notice that for the values of  $N_f=9,15$  shown in Fig.~\ref{betaQCDMS}, the 4 and 5 loop MG results are in good agreement with each other and therefore all the fixed points found are presumably reliable. Also, for $N_f=15$, both 4 and 5 loop MG predictions practically coincide with the perturbation theory, as expected, since the corrections to the perturbative expansion are small.

Variety of other approaches exist on the market to estimate the size of the IR conformal window among which are numerical lattice simulations \cite{Lombardo:2014mda}, the analytic solutions to Schwinger-Dyson equations \cite{Appelquist:1996dq}, functional RG method\cite{Braun:2010qs}, estimates based on the conjectured form of the all-order QCD beta function\cite{Antipin:2009wr,Pica:2010mt}, and holographic models\cite{Jarvinen:2011qe}. These approaches produced a variety of predictions for $N_f^{IR}$ ranging, between 7 and 12.

\section{Conclusions}\label{Sec:end}

In this paper, we have made an attempt to approach the non-perturbative regime of a QFT using the state-of-the-art perturbative expressions.
We used the recent method proposed in Ref.~\cite{Mera:2018qte}, which exploits the Meijer G-functions in order to resum divergent series knowing the first few terms of their perturbative power series expansion.
Specifically, we focused on the $\beta-$functions of the $\phi^4$ model and $U(1),SU(3)$ gauge theories assuming the $\beta-$functions of these QFTs to be analytic which is fundamental for the applicability of MG approach. For the singular functions, the usual Borel-Pad\'e approximation is still better as stressed in Ref.~\cite{Mera:2018qte} and, for example, this would be the case of supersymmetric gauge theories whose $\beta-$functions have simple poles~\cite{Shifman:1986zi}.

For the $\phi^4$ model, we used the MG method on the perturbative $\beta-$function up to 7-loop and found that no fixed point emerges for any value of the quartic coupling constant which is in agreement with the results known in the literature. Remarkably, we also found that precisely for MG on the 7-loop $\beta-$function, the result agrees with the asymptotic behavior computed using the semiclassical approach.

For the gauge theories, the $\beta-$function is known up to 5-loop and, to check the convergence of the MG approximants, we estimated higher order corrections from the large $N_f$ expansion and then extrapolated to lower $N_f$. First of all, we noticed that the convergence of the MG algorithm is not very good, but still qualitatively sufficient to
infer physical information. In particular, for $U(1)$ case the algorithm established the non-existence of fixed points below the renormalon constraint and we argued that this conclusion holds for any value of the number of flavors $N_f$. For $SU(3)$ gauge theory the situation is richer as the number of flavors varies. We found that from $N_f=0$ up to $N_f=9$, no fixed points exist using the known 5-loop beta function. Starting from $N_f=9 $ an IR, non-perturbative fixed point develops which becomes perturbative for $N_f \approx 16$ and finally disappears for $N_f=16.5$. For a larger number of flavors, a UV fixed point of the beta function was found which first develops for $N_f\approx30$
and which, for $N_f\approx 58$, reaches the renormalon bound.

Furthermore, a possible consequence of this resummation method is that also the renormalon series might be resummable, even though a clarification of the non-perturbative ambiguity is still lacking. However, such an ambiguity is not specific to the problem of renormalons and affects the method per s\'e, depending on the series under consideration. In any case, it is interesting to inquire how this possible renormalon summation could affect the non-perturbative physics. Partial answers may be extrapolated from the QCD literature, in which the renormalon divergence is used to estimate the non-perturbative power corrections to the heavy quark masses~\cite{Beneke:1994sw,Bigi:1994em}.

Another consequence is to see whether the MG algorithm can give some information about the non-perturbative mechanism for the quark confinement. To this aim, see for example  Refs.~\cite{Beneke:1992ch,Aglietti:1995tg,Beneke:1998rk,Hoang:1998nz}, where the usual ambiguity in the Laplace transform is used to estimate the non-perturbative power corrections to the quark anti-quark potential. It would be interesting to see whether the MG algorithm can give a more precise estimate on these issues. Another possible and intriguing question would be how to approach the multi-coupling resummation problem within this method, and even more specifically how to attack the generalized renormalons in multi-variables~\cite{Maiezza:2018pkk}.

%
%
\section*{Acknowledgements}

We thank Bla\v zenka Meli\'c and Hector Mera for a careful reading of the manuscript and valuable suggestions.
OA was partially supported by the Croatian Science Foundation project number 4418. OA and AM acknowledge partial support by the H2020 CSA Twinning project No. 692194, ''RBI-T-WINNING''. JV was funded by  Fondecyt project N. 3170154 and partially supported by  Conicyt  PIA/Basal FB0821.

%
%

\appendix

\section{The Borel-hypergeometric  resummation procedure for divergent series}\label{App_MeijerG}

 \begin{figure}\centering
     \includegraphics[scale=0.3]{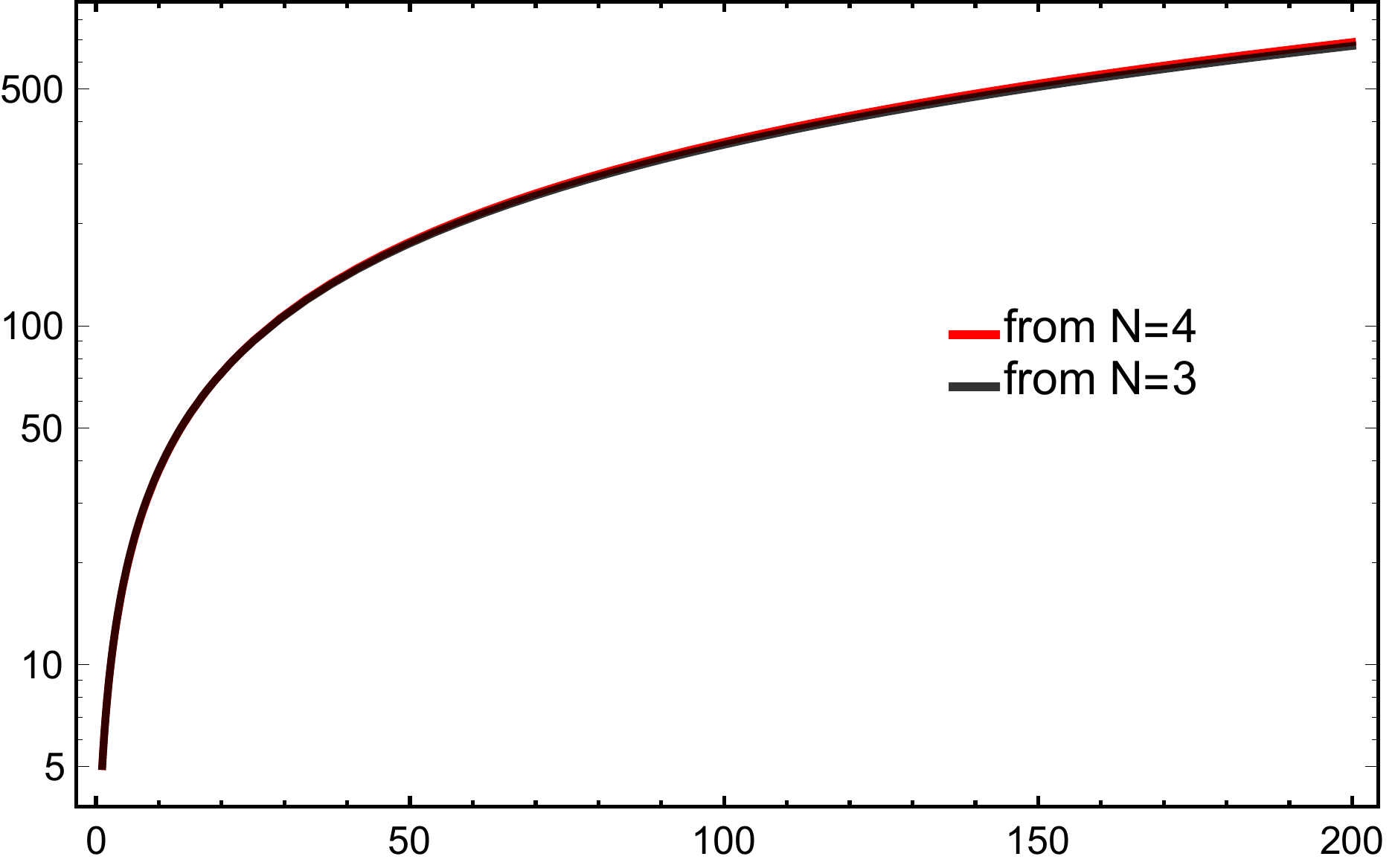}
  \caption{Example of a comparison of the Meijer G-function approximant with $N=3,4$, with the coefficients $r,s,w,z$ satisfying~\eqref{condition}.}
  \label{fig:comparison}
\end{figure}

\subsection{Constructing the algorithm}

First of all, we recall the algorithm as proposed in~\cite{Mera:2018qte}, which we have used in the main text, and afterward we add further explanations on its subtle points. Finally, we also discuss the convergence of the algorithm without pretending to perform a full analysis on the subject.

\paragraph{The Algorithm.}\label{algo} Let us consider a partial
sum of power series in variable $g$ with coefficients $a_1,a_2...a_N$ and assume for the moment $N$ to be odd-number. The algorithm goes as follows:
\begin{enumerate}
\item One computes the Borel-transformed coefficients $b_n=a_n/n!$ and makes the ansatz that the ratio of the consecutive Borel-transformed coefficients is a rational function of $n$
\begin{equation}\label{rational}
\frac{b_{n+1}}{b_n}=r_N(n)=\frac{\sum_{m=0}^{l}p_m n^m}{1+\sum_{m=1}^{l}q_m n^m}\,,
\end{equation}
with $l=(N-1)/2$ and $n$ ranges from 0 to $N-1$. Thus, we have $N$ unknowns $p_m,q_m$ which are determined from $N$ equations in~\eqref{rational}.

\item The heart of the algorithm in~\cite{Mera:2018qte} is to find the hypergeometric vectors $\overline{x}=(1,-x_1,...,-x_l)$, $\overline{y}=(-y_1,...,-y_l)$
via the equations
\begin{eqnarray}\label{heart}
\sum_{m=0}^{l} p_m x^m &=& 0 \\ \nonumber
1+\sum_{m=1}^{l} q_m y^m &=& 0\,,
\end{eqnarray}
and then define the hypergeometric Borel approximant(s) through the generalized hypergeometric function
\begin{equation}\label{Hyper}
B_N(z)= {}_{l+1}F_l (\overline{x},\overline{y},\frac{p_l}{q_l}z)\,.
\end{equation}

\item Finally, one has to go back to the original series in the variable $g$ and remove the $1/n!$ via the Laplace transform of~\eqref{Hyper},
ending up with the resummed  series $S(g)$, which can be represented in the form
\begin{equation}\label{MJG}
S(g)=\frac{\prod_{i=1}^l \Gamma(-y_i)}{\prod_{i=1}^l \Gamma(-x_i)} G_{l+1,l+2}^{l+2,1}
\left(
\begin{array}{c}
1,-y_1,...,-y_l \\
1,1,-x_1,...,-x_l
\end{array}
\right.
\left|  \left. -\frac{q_l}{p_l g}\right. \right)  \,,
\end{equation}
where $G$ stands for the Meijer G-function and $\Gamma$ is the Euler's Gamma function.
\end{enumerate}
The generalization to an even $N$ is straightforward. It is enough to subtract from the original series the constant term, then factor out the linear term and then apply the algorithm for odd $N$ as above. Finally, one has to re-multiply the result with the linear term and re-add the constant. An explicit example for $N=3$ vs $N=4$ will be given below.

As discussed in Subsec.~\ref{Subsec:ren}, in order for the MG algorithm to work, it is crucial that the perturbative expansion matches the full answer in the limit of very small coupling. In some cases, the exact answer might be known while in the others, such as QED, one simply assumes that such matching is justified based on the comparison with experimental measurements. In any case, upon this matching the real part of the output is meaningful and the possible appearance of an imaginary part cannot be understood in terms of MG alone. One needs, for example, to resum  the trans-series built from the perturbative problem in terms of MG, to prove that the imaginary part vanishes while the real one of the original MG answer converges to the true result.

\paragraph{More insights.} The above algorithm makes use of the Pad\'e approximation in~\eqref{rational}, the hypergeometric approximation~\cite{Mera:2014sfa} in~\eqref{Hyper} and in the final step approximation~\eqref{MJG} via the Meijer G-function. The last two steps exploit the ''flexibility'' of
the corresponding functions, i.e. the feature of containing most of the regular and special functions as a particular case. Perhaps a less clear step
is the one in~\eqref{heart} so let us go through it in detail. The hypergeometric function in~\eqref{Hyper}
can be Taylor expanded in terms of Pocchammer symbols and we denote with $H(n)$ the $n-$th term of such series. The Eq.~\eqref{heart} are then exactly equivalent
to solve
\begin{eqnarray}\label{understand}
  \lim_{n\rightarrow\infty} \frac{H(n+1)}{H(n)} &=& \lim_{n\rightarrow\infty}r_N(n) \\ \nonumber
  \frac{H(n+1)}{H(n)} &=& r_N(n)   \,\,\,\,\,\,\,\,\,\forall n \in[0,N-1] \,,
\end{eqnarray}
which can be easily verified directly, for example for $N=3$. Eq.~\eqref{understand} makes more manifest the meaning
of~\eqref{heart}: one imposes an asymptotic condition in the first equation of~\eqref{understand} plus the three further equations consistent with~\eqref{rational},
for a total of four unknowns, completely determining the hypergeometric Borel approximant (for $N=3$) as well as the final Meijer G-function approximant $S(g)$.

\subsection{Convergence}

Since this resummation method based on Meijer G-function is new, we do not know about its convergence and this was also pointed out in Ref.~\cite{Mera:2018qte}.
In this subsection we focus on the convergence of the algorithm at order $N=3$ vs order $N=4$. For this comparison, consider the partial sum
\begin{equation}\label{fewnominal}
\tilde{S}(g)=1+r g+s g^2+ w g^3 + z g^4\,,
\end{equation}
with $r,s,w,z$ being arbitrary coefficients. Since we want to compare the algorithm at orders $N=3,4$, first we apply the above algorithm for odd $N=3$ truncating
the $\tilde{S}(g)$ at order $g^3$. The result can be Taylor expanded as
\begin{equation}\label{prediction}
1+r g+s g^2+ w g^3 +\frac{2 g^4 w \left(8 r^2 w-3 r s^2-3 s w\right)}{9 r^2 s+r w-6 s^2}+\mathcal{O}(g^5)\,,
\end{equation}
thus one has a \emph{prediction} for the coefficient $z$. On the other hand, one can employ the algorithm on the full $\tilde{S}(g)$, now with even $N=4$, by rewriting~\eqref{fewnominal}
as
\begin{equation}\label{fewnominal1}
\tilde{S}(g)=1+r g(s/r g+ w/r g^2 + z/r g^3)\,,
\end{equation}
then applying the odd-$N$ algorithm on $(s/r g+ w/r g^2 + z/r g^3)$, multiplying the result by $r g$ and, finally, adding the constant $1$ to the final result. For consistency, the Taylor expansion of the latter up to $g^{4}$ is trivially identical to~\eqref{fewnominal1}. With both results for $N=3,4$ at hand, one can compare them
to see when the $N=3$ algorithm converges to the $N=4$ one. Now the point is that assuming the predictivity of the leading algorithm to determine the coefficient $z$, one gets
the condition on $r,s,w,z$
\begin{equation}\label{condition}
z=\frac{2 w \left(8 r^2 w-3 r s^2-3 s w\right)}{9 r^2 s+r w-6 s^2}\,.
\end{equation}
This condition is sufficient because, starting from a fewnominal order $3$, through the algorithm order $N=3$ one builds a fewnominal order $4$ and thereby applies the algorithm for $N=4$. This is also a posteriori checked numerically in Fig.~\ref{fig:comparison} for a given solution of~\eqref{condition}. Curiously, the condition in~\eqref{condition} is not necessary, albeit sufficient. This can be easily shown by a direct numerical counter-example, finding a combination of $r,s,w,z$ that violates~\eqref{condition} but yields convergence. The reason is that one can follow again the same logic leading to Eq.~\eqref{condition} and find similar conditions by considering higher coefficients coming from the Taylor expansions of the MG outputs for both $N=3,4$. Then, it might happen that even though Eq.~\eqref{condition} is violated, the conditions for convergence between $N=3$ and $N=4$ algorithms from the higher order terms are satisfied. As a qualitative recipe, it is worth noticing that condition~\eqref{condition} works roughly even perturbing the equality up to $\approx50\%$.


\bibliographystyle{jhep}
\bibliography{biblio}

\end{document}